\documentclass[10pt, preprint2]{aastex}

\newcommand{\msun}{\mathrm{M_{\odot}}}

\begin{document}

\shortauthors{Walker \& Lewis}
\shorttitle{Nanolensing of GRBs}

\title{Nanolensing of gamma-ray bursts}

\author{Mark A. Walker$^{1,2,3}$, Geraint F. Lewis$^{1,4}$}

\affil{1. School of Physics, University of Sydney, NSW 2006, Australia}
\affil{2. Australia Telescope National Facility, CSIRO, Epping 1710, Australia}
\affil{3. Raman Research Institute, Bangalore 560012, India}
\affil{4. Anglo-Australian Observatory, PO Box 296, Epping 1710, Australia}

\begin{abstract}
All quasars vary  in their optical flux on a  time-scale of years, and
it  has been  proposed that  these variations  are principally  due to
gravitational lensing by  a cosmologically distributed population
of planetary  mass objects.  This interpretation has  implications for
the observable  properties of gamma-ray  bursts (GRBs) -- as  a source
expands  across  the nano-arcsecond  caustic  network, variability  is
expected -- and data on GRBs can be used to test the proposed model of
quasar variability. Here we employ an ultra-relativistic blast-wave
model of the source, with no intrinsic variations, to study
the effects of nanolensing on GRBs. Taken in isolation the light-curves
of the caustic crossings are predictable, and we find that a subset of
the predicted light-curves (the image-annihilating  fold crossings)
resemble the ``pulses'' which are commonly seen in long GRBs. Furthermore,
for sources at high redshift the expected time between  caustic crossings
is  of order  seconds,  comparable to  the observed time  between pulses.
These  points suggest that it might be possible to model some
of the observed variations of GRBs in terms of  nanolensing; however,
our simulated light-curves  exhibit a small depth of  modulation compared
to what is observed. This means that the GRB data do not significantly
constrain the quasar nanolensing model; it also means that the simplest
nanolensing model cannot explain the observed GRB ``pulses''. Viable 
nanolensing models for pulses probably require a large external beam
shear.  If  a  viable model  can  be constructed it would effect  a
considerable simplification in source modelling and,  ironically, it
would explain why  no macro-lensed GRBs have been identified to date.

Independent of the particular theoretical model, we can test for
the presence of nanolensing in GRB data because
any  variability  due to  nanolensing  should  manifest parallax:  the
timing of  caustic crossings, and  hence the temporal  substructure of
bursts should  be different as seen by  separated observers.  Parallax
therefore  shifts triangulated  burst locations  away from  their true
positions; this  displacement is typically  expected to be at  the few
arcminute level, and existing astrometry  is not good enough to reveal
the predicted  effects.  Useful constraints can,  however, be obtained
by  comparing   the  relative  timing  of  individual   peaks  in  the
light-curves  recorded by spacecraft  in the  Inter-Planetary Network;
published data  show hints of  the predicted temporal shifts,  but the
photon counting statistics are not good enough to categorically decide
the matter. There is  no plausible alternative interpretation for this
phenomenon, and  if it is confirmed  as a real effect  then it compels
acceptance of  a cosmology that  is very different from  the currently
popular model.
\end{abstract}


\section{Introduction}
At present  we do not know  what constitutes the bulk  of the material
Universe, i.e. the dark matter.  Many dark matter candidates have been
proposed,  ranging from  elementary particles  to  macroscopic objects
such as black holes --  see, e.g., Trimble (1987), Ashman (1992), Carr
(1994). To  date it  has been possible  to eliminate  some suggestions
(e.g.   brown dwarfs:  Tinney 1999),  by  virtue of  a clear  conflict
between models  and data, but  a positive identification has  not been
achieved. One proposal whose implications have not yet been thoroughly
explored is that of Hawkins (1993,  1996), who argued, on the basis of
photometric  monitoring  of  quasars,  that the  Universe  contains  a
near-critical density  of planetary-mass  objects (cf. Press  and Gunn
1973).   Gravitational lensing  by  such a  population  is capable  of
explaining much  of the observed  variability (see also  Schneider and
Wei\ss\ 1987;  Schneider 1993) --- though  this should not  be taken to
imply  that intrinsic  variations are  absent.  This  suggestion  is a
radical departure from the now-standard picture
of  a  universe  dominated  by  elementary  particles
(i.e. Cold  Dark Matter -- see,  e.g., Peebles 1993;  Blumenthal et al
1984;  Davis  et  al   1985),  and  has  received  little  theoretical
attention.   Building on  earlier  work (Schneider  and Wagoner  1987;
Rauch 1991; Seljak and Holz 1999;  Metcalf and Silk 1999), 
Minty, Heavens  and Hawkins
(2001) described  a  statistical  test  of  the  model  based  on  the
light-curves  of distant supernovae,  but existing  data do  not allow
these  ideas to  be  usefully  implemented. A  test  based on  surface
brightness variability in low-redshift  galaxies has been described by
Lewis and Ibata (2001), but this idea also requires data which are not
yet available, as does the test based on monitoring quasars seen through
low-redshift clusters of galaxies (Walker and Ireland 1995; Tadros,
Warren and Hewett 1998). Data from the MACHO and EROS experiments exclude
planetary-mass compact objects as a significant contributor to the
Galactic dark matter (Alcock et al 1998). However objects which are
sufficiently compact that they qualify as strong gravitational lenses
at cosmological distances are not necessarily strong gravitational
lenses when they are located in the Galactic halo (Walker 1999;
see also Draine 1998, Rafikov and Draine 2001). The Galactic
microlensing experiments therefore do not directly test the quasar
nanolensing hypothesis.

It  is  obvious  that  one  could  attempt  to  devise  further,  more
sophisticated tests  of the nanolensing interpretation based  on existing
quasar data, but it is a priori unlikely that any such test will yield
a definitive result. The main reason  for this is source size: the low
amplitude of variations seen in quasar optical light-curves means that
these  sources  are necessarily  larger  than  the  scale-size of  the
hypothesised  caustic  structure,  thus  smoothing  the  magnification
pattern,  and in  the  process destroying  the high  spatial-frequency
information. (This would be less of  a barrier to progress if we had a
reliable model for the physical structure of quasars, so that detailed
predictions  for their  lensed  appearance could  be  given with  some
confidence.) For  example the approach  of Dalcanton et al  (1994; see
also Canizares 1982), based on  the statistics of equivalent widths of
quasar emission lines,  is quite insensitive in this  regime where the
continuum  source is  resolved  by  the lenses.  To  make progress  we
therefore  require  a numerous  population  of small,  highly-luminous
sources  at  large distances;  gamma-ray  bursters  constitute such  a
population. Gravitational  lensing of gamma-ray  bursts has previously
been  considered  by a  number  of  authors  (e.g. Paczy\'nski  1986b;
McBreen and  Metcalfe 1988;  Mao 1992; Gould  1992; Blaes  and Webster
1992; Nemiroff and Gould  1995), but their considerations are relevant
to lenses  which are  either much more  massive, or much  less massive
than the planetary-mass range which we consider here. The present work
also differs  substantively from  previous investigations in  that the
evolution (expansion) of the source structure plays a critical role
in our analysis.

Following  the  launch of  the  Compton  Gamma-Ray Observatory,  BATSE
(Burst And Transient Source Experiment) soon discovered that gamma-ray
bursts are isotropically distributed on  the sky, and that they depart
from Euclidean source counts at  low flux levels (Fishman et al 1994).
These   discoveries   immediately    shifted   attention   away   from
interpretations based on Galactic  neutron stars, which had previously
been popular, towards a variety  of models in which energies amounting
to  a significant  fraction of  $\msun  c^2$ are  rapidly released  by
sources at  cosmological distances (see  Paczy\'nski 1995).  Discovery
of X-ray afterglows from these  transient events (Costa et al 1997) by
BeppoSAX allowed the sources to be accurately positioned, and thus led
to the discovery  of optical and radio afterglows  (van~Paradijs et al
1997; Djorgovski et al 1997;  Frail et al 1997). Spectroscopy of these
optical transients  has in some  cases revealed absorption  lines from
gas  at redshifts  $z_s\ga1$,  thus firmly  establishing the  distance
scale  of  the  bursters  to  be cosmological  (Metzger  et  al  1997;
van~Paradijs, Kouveliotou and Wijers 2000).

Although we still  do not know what process injects the 
burst energy, it is broadly agreed that the   radiation  observed
from GRBs arises  from  a  relativistically  expanding
source.  Relativistic  expansion is required  in order that  we should
see  gamma-rays at  all  --  at least  at  energies $\ga$~MeV  --
otherwise the inferred source-size is  so small, and the photon energy
density so high,  that two-photon pair-production converts essentially
all  the  gammas  into  material  particles (Cavallo  and  Rees  1978;
Fenimore,  Epstein  and  Ho  1993;  Baring and  Harding  1997).  These
considerations  require expansion  with  Lorentz factors  of at  least
$\Gamma\sim10^2$.

At  an  early  stage  in  the  modelling  of  gamma-ray  bursts  in  a
cosmological  context,  it was  pointed  out  that gamma-ray  emission
should arise as the the  ambient medium undergoes shock compression by
the expanding  material (Rees \&  M\'esz\'aros 1992). This  picture --
the ``blast-wave'',  or ``external shock''  model -- now  provides the
accepted  context for  modelling  the broad-band  (X-ray, optical  and
radio) afterglows of bursts (Paczy\'nski and Rhoads 1993; M\'esz\'aros
and  Rees 1997;  Granot, Piran  and Sari  1999), but  has  fallen into
disfavour  as an  interpretation of  the prompt  gamma-ray  burst. The
reason  for the  demise of  this model  is simply  that it  cannot, in
itself,  accommodate the rapid,  large-amplitude variability  which is
manifest in  the gamma-ray data (Fenimore, Madras  and Nayashkin 1996;
Sari  and Piran  1997;  see  also Dermer  and  Mitman 1999;  Fenimore,
Ramirez-Ruiz and Wu 1999).  This  point has promoted the idea that the
gamma-rays arise from internal shocks within the relativistic outflow,
so that the  temporal variations of the burst  reflect the input power
variations of the source  (e.g. Rees and M\'esz\'aros 1994; Kobayashi,
Piran and Sari 1997).

Because  we   are  investigating   an hypothesis  in   which  apparent
variability   arises  external  to   the  source,   as  a   result  of
gravitational  lensing,  arguments  which  assume  that  the  observed
variations are  intrinsic immediately lose  most, if not all  of their
force. Given  this there are at  least three motivations  to return to
the blast-wave interpretation of the prompt gamma-ray emission: first,
this picture allows the initial  deposition of energy to be impulsive,
and  is  therefore  suitable  for  a  broad  range  of  source  models
(independent  of the  specific physics  of the  energy input)  with no
fine-tuning  required;  secondly,  the  low  radiative  efficiency  of
internal shocks  (Spada, Panaitescu and M\'esz\'aros  2000) appears to
be inconsistent  with the event  energetics in at least  some cases
(Paczy\'nski 2001); and thirdly,  gamma-ray emission from  an external
shock  ought  to be  present  at some  level.   This  emission can  be
sensibly described  by a self-similar,  decelerating blast-wave model,
similar to  those developed for  the afterglow emission  (e.g. Granot,
Piran  and Sari  1999), but  radiative rather  than adiabatic.  Such a
description leads  us to  expect a circular  source with  {\it very\/}
strong limb brightening and consequently we   anticipate   significant
flux variations if  the limb crosses a caustic.  This thin, bright and
rapidly expanding ring is  a near-perfect instrument for revealing any
caustic structure that might be present along the line-of-sight to the
source --- see also Loeb and Perna (1998), and Mao and Loeb (2001), who
considered microlensing of optical afterglows by stellar-mass objects
in the low optical-depth limit.

In this  paper we use the  source model just described  to explore the
hypothesis that the Universe contains a high density of planetary-mass
objects,  with an  optical  depth to  gravitational  lensing of  order
unity.  Often  the characteristic  angular  scale  of  the lenses  (in
arcseconds)  is  indicated in  the  nomenclature  given to  associated
phenomena (e.g.  {\it micro\/}lensing), and  following this convention
dictates the  name ``nanolensing''  for the effects  discussed herein.
Although  gravitational lensing  is the  main focus  of  our attention
here, in  studying the  variations which might  be introduced  by this
process we are  also implicitly addressing the physics  of the sources
themselves. We start by presenting our source model in \S2, and in \S3
we turn  to aspects of  the gravitational lensing;  light-curves which
result from the marriage of  these elements are presented and compared
with data in \S4. The role of parallax is explored in \S5, followed by
a  discussion of  related issues  concerning both  the lenses  and the
sources (\S6), and our conclusions are given in \S7.

\section{Source model}
\label{sec:sourcemodel}
Following    Rees    and    M\'esz\'aros    (1992),   we    adopt    a
spherically-symmetric,  ultra-relativistic  blast-wave  model  of  the
gamma-ray   burst   phenomenon,   in   which   a   total   energy   of
$10^{52}\,E_{52}\;{\rm erg}$  resides in  ejecta which expand  with an
initial Lorentz factor $\gamma=10^3\gamma_3\ga10^3$ into a homogeneous
medium  of  density $n\;{\rm  cm^{-3}}$.  At  first  the ejecta  coast
($\gamma\simeq$~constant),  but this  lasts  only for  a brief  period
$\sim0.1\;\gamma_3^{-8/3}(E_{52}/n)^{1/3}\;{\rm  s}$, and subsequently
the  ejecta decelerate, with  some fraction  of the  thermalised power
appearing as radiation. We  will treat these as distinct, self-similar
phases of evolution, characterised by the index $m$, where
\begin{equation}
\Gamma^2=\Gamma_0^2\left({{R_0}\over{R}}\right)^m
\end{equation}
at radius  $R$, and $\Gamma=\sqrt{2}\gamma$  is the Lorentz  factor of
the  shock-wave  which  precedes   the  ejecta  (Blandford  and  McKee
1976). In the coasting phase  we evidently have $m=0$; while $m=3$ for
adiabatic evolution,  and $m=12$ for  a fully radiative  blast-wave --
i.e. one in  which all of the thermalised  energy is promptly radiated
away  (Cohen, Piran and  Sari 1998).   The adiabatic  approximation is
appropriate for  the afterglow, when the radiation  time-scale is much
longer than  the expansion time-scale; for the  gamma-ray burst itself
we adopt  the fully radiative  solution, $m=12$. In  this circumstance
the gamma-rays  arise from a  thin shell immediately behind  the shock
front.

It is convenient  to take $R_0$ as the radius  at which the transition
between  coasting and  semi-radiative  solutions takes  place, and  to
model  the  evolution   of  the  blast  wave  as   if  there  were  an
instantaneous transition between these solutions, as the shock crosses
this radius.   We can estimate the  value of $R_0$ by  noting that the
shock  will start  to decelerate  when a  significant fraction  of the
initial  energy   of  the  blast   has  been  thermalised   (Rees  and
M\'esz\'aros 1992):
\begin{equation}
R_0\simeq1.2\times10^{16}\left({{E_{52}}\over{n\Gamma_3^2}}\right)^{1/3}
\qquad{\rm cm}.
\end{equation}
Where   numerical  estimates   are  required   we  adopt   the  values
$E_{52}=\Gamma_3=n=1$ throughout. For some  estimates we need to know,
in addition,  the Hubble constant,  which we take to  be $H_0=70\;{\rm
km\,s^{-1}\,Mpc^{-1}}$.

In  this paper we  consider only  bolometric radiation  properties, so
that it is not necessary to specify the radiation mechanism.

\subsection{Kinematics}
Because  the expansion is  spherically-symmetric, the  observed source
structure is axisymmetric,  and at any given time  it can be expressed
as a function of a single  variable, such as the apparent radius, $r$,
as measured  by a distant  observer, relative to the  line-of-sight to
$R=0$.  Because the  expansion  is relativistic,  it  is essential  to
incorporate light travel-time in  any computation of the appearance of
the source. In terms of the  time, $t$, measured by a distant observer
{\it at the same redshift\/} (with $t=0$ corresponding to the start of
the expansion),  there is a  maximum apparent radius,  $r_{max}$, from
which photons can be received by the observer:
\begin{equation}
r_{max}^{2m+2}=\Gamma_0^2R_0^m
\left[\left({{2m+2}\over{m+2}}\right)ct\right]^{m+2},
\end{equation}
and this  defines the limb  of the source  at any time. In  general we
expect the source to be at  non-zero redshift, $z_s$, and to allow for
this one  simply makes the replacement  $t\rightarrow t/(1+z_s)$, both
here and in subsequent formulae. In cases where a
numerical estimate is required,  we adopt the value $z_s=5$ throughout
this paper; this corresponds to  the median redshift in the GRB source
population model of  Bromm and Loeb (2002). The  radius at $r=r_{max}$
is given by $R_{max}$:
\begin{equation}
R_{max}^{m+1}=2ct\,\Gamma_0^2 R_0^m\left({{m+1}\over{m+2}}\right),
\end{equation}
and  if we  define  $q\equiv R/R_{max}$,  then  for any  point on  the
emitting surface, at any given time, we have
\begin{equation}
\tilde r^2={q\over{m+1}}\left[(m+2)-q^{m+1}\right],
\end{equation}
where $\tilde r\equiv r/r_{max}$. Notice  that there are two values of
$q$ which correspond  to each value of $\tilde  r$ (other than $\tilde
r=1$); one  of these values  lies in the  range $0\le q<1$,  while the
other lies in the range $1<q\le Q$, with $Q^{m+1}=m+2$.

Knowing the apparent size of the  source as a function of time, we can
easily find  the apparent expansion  speed, $\beta_{ap}$ (in  units of
$c$), just by differentiating equation 3:
\begin{equation}
\beta_{ap}^{2m+2}=\Gamma_0^2 \left[{{(m+2)R_0}\over{(2m+2)ct}}\right]^m.
\end{equation}

\subsection{Intensity profile}
By  virtue  of being  almost  coincident  with  the shock  front,  the
emitting surface exhibits the geometry  of a shell moving with Lorentz
factor  $\Gamma$,  as  described  in  \S2.1.   However,  the  emitting
particles are part of the post-shock flow so their bulk Lorentz factor
is $\gamma=\Gamma/\sqrt{2}$, and in consequence the limb of the source
($\tilde  r=1$)  does not  correspond  to  the  peak of  the  observed
intensity.   This  point  does  not  appear to  have  been  recognised
previously (cf. Granot and Loeb  2001; Gaudi, Granot and Loeb 2001).
For a thin emitting shell, the
peak surface  brightness corresponds  to polar angle  $\psi=\pi/2$ (as
measured in the co-moving frame)  relative to the surface normal.  For
an  emitting shell  of fractional  radial thickness  $\Delta\ll1$, the
bolometric surface brightness can be described by
\begin{equation}
I(r,t)={{{\cal D}^4\,{\cal I}}\over{\sqrt{\cos^2\psi + \Delta/4}}},
\end{equation}
with ${\cal I}$  being the bolometric intensity emitted  normal to the
surface,  as measured in  the rest  frame of  the emitting  shell, and
${\cal  D}$  the  Doppler   factor.  (Recall  that  $q(\tilde  r)$  is
double-valued, so  that there are implicitly two  contributions to the
right-hand-side of equation 7.)  It is easy to show that rays reaching
the observer at a given instant satisfy
\begin{equation}
\cos\psi={{(2m+3)q^{m+1}-(m+2)}\over{(2m+1)q^{m+1}+(m+2)}},
\end{equation}
and that the ratio of observed-to-emitted photon energies is
\begin{equation}
{\cal D}={{4(m+1)\gamma_{max}q^{(m+2)/2}}\over{(2m+1)q^{m+1}+(m+2)}},
\end{equation}
where  $\gamma_{max}\sqrt{2}:=\Gamma_0 (R_0/R_{max})^{m/2}$.   Now the
total    energy   density    in   the    post-shock   gas    is   just
$4\gamma^2n\,m_pc^2$ (Blandford  and McKee 1976), and in  the frame of
this material the shock itself moves at speed $c/3$, so if all of this
power  is promptly radiated,  the emergent  normal intensity  (i.e. at
$\psi=0$) is
\begin{equation}
{\cal I}\simeq{1\over{3\pi}}\gamma^2n\,m_pc^3.
\end{equation}
The  above  set of  equations  provides  a  complete prescription  for
calculation of  the bolometric intensity profile;  the fully radiative
case, corresponding to  $m=12$, is shown in figure  1 for a fractional
shell thickness of  $\Delta=4\times10^{-3}$.  This choice for $\Delta$
is appropriate to, for  example, synchrotron emission at $\sim200$~keV
if  the magnetic field  energy-density is  10\%\ of  the equipartition
value.   The essential  feature of  this profile  is that  it  is {\it
very\/}   strongly    limb-brightened,   with   a    sharp   peak   at
$r\simeq0.99295\,r_{max}$  (corresponding  to $q^{m+1}:=(m+2)/(2m+3)$,
i.e.  $\psi=\pi/2$). Source  intensity  profiles for  any index  $3\le
m\le12$ exhibit this limb-brightening effect, but as $m$ increases the
limb-to-centre     intensity      ratio     grows     according     to
$(32/27)(m+2)^{3m/(m+1)}$,  and  the  peak-to-centre  intensity  ratio
equals  $(2m+3)^{3m/(m+1)}/8\sqrt{\Delta}$; the  peak  also shifts  to
larger radii with increasing $m$.

The received flux is computed in the usual way, i.e.
\begin{equation}
F=\int\!\!{\rm d}\Omega\,I\quad\propto\quad2\pi r_{max}^2\int_0^1\!\!
{\rm d}\tilde r\,\tilde r\,I.
\end{equation}
Using equation  5 it is  straightforward to rewrite the  integral over
$\tilde r$ as  an integral over $q$ such that $0\le  q\le Q$.  In this
way  we  find  that  the   observed  flux  should  vary  as  $F\propto
t^{(2-m)/(m+1)}$,  so  that  $F$  grows  as $t^2$  during  the  early,
coasting phase  ($m=0$), and then declines as  $t^{-10/13}$ during the
self-similar radiative ($m=12$)  evolution. A simple model light-curve
can  thus be  constructed by  applying  these two  solutions in  their
appropriate  regimes (small/large  $t$,  respectively), and  switching
between the two  at the point where they cross.  That is the procedure
we  shall  follow  here  ---  our  calculations  are  intended  to  be
illustrative and high precision is not needed.  One important point to
note  is that the  luminosity in  the self-similar  deceleration phase
declines too slowly  to yield a finite radiated  energy at late times,
and so the model clearly has a limited range of applicability.


\begin{figure*}
\plotone{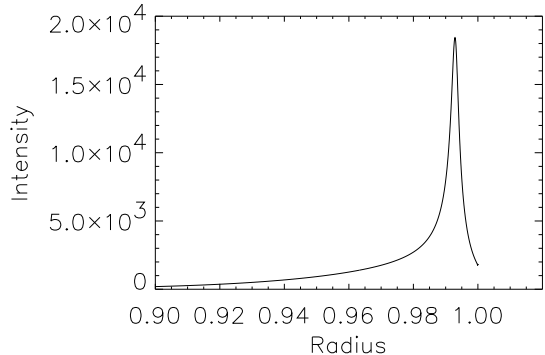}
\caption{The  radial  distribution  of  bolometric  intensity,  for  a
blast-wave  in self-similar,  fully-radiative ($m=12$)  evolution. The
emitting  shell  is assumed  to  have  a  fractional radial  thickness
$\Delta=4\times10^{-3}$.  The intensity is  normalised to  the central
intensity, and declines monotonically beyond the left-hand edge of the
plot.  The location of the peak ($\tilde r\simeq0.993$) corresponds to
$\psi=\pi/2$, i.e. tangent to the emitting shell.}
\end{figure*}

\begin{figure*}
\plotone{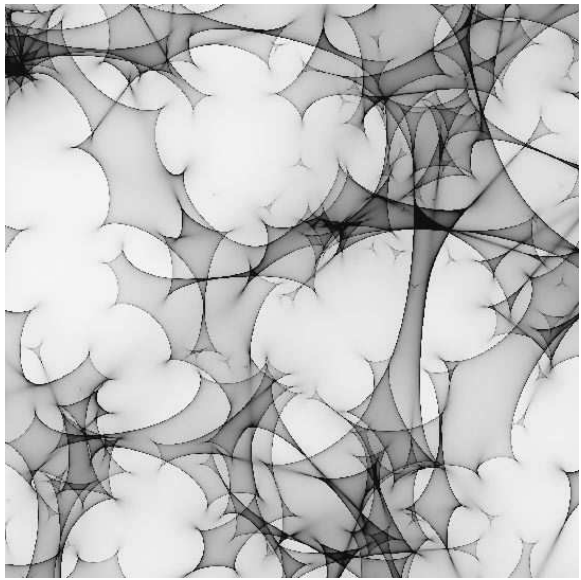}
\caption{One realisation  of a magnification pattern  appropriate to a
source at  redshift $z_s=5$, in  a universe populated  with point-mass
lenses  such that  $\Omega_{lens}=\Omega=1$. This  pattern  is derived
from  a simulation in  which the  locations of  the lenses  are chosen
randomly within a  single plane, having a total  optical depth of 0.6.
The rays are chosen to have  a uniform density across the image plane;
this figure  shows their density  across the source plane;  regions of
high  ray  density  in  this   figure  are  therefore  areas  of  high
magnification, and most of  these regions correspond to fold caustics.
Our  simulations  utilised code  from  Wambsganss,  Paczy\'nski \&  Katz
(1990).}
\end{figure*}

\begin{figure*}
\plotone{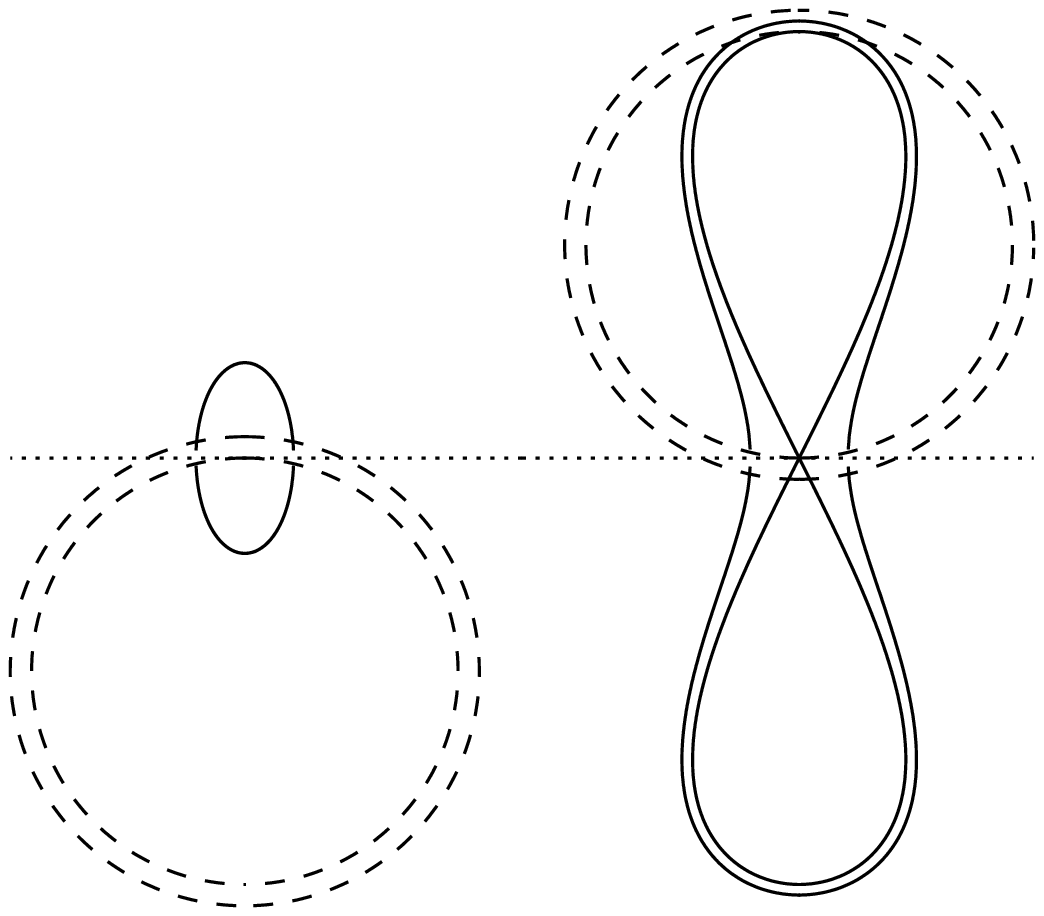}  \figcaption{The  images (solid  lines)  of two  thin
(10\% fractional  width), annular  sources (dashed curves),  on either
side of  a fold  caustic (dotted line).  For convenience  the critical
curve ($X=0$) has been superimposed  on the caustic ($x=0$). The upper
half of this  figure is the three-image region, and  the lower half is
the  single-image region.  In  each  case the  inner  boundary of  the
annulus just  touches the  caustic, and the  centre of each  source is
assumed to lie at a distance of $s=\pm1/a$ from the fold.}
\end{figure*}


\section{Lens model}
\label{sec:lensmodel}
If the Universe contains a mean mass density in lenses $\Omega_{lens}$
(in  units  of  the  critical  density), then  the  optical  depth  to
gravitational lensing is  $\tau\simeq0.11\,\Omega_{lens}$ for a source
at redshift $z_s=1$,  and $\tau\simeq0.62\,\Omega_{lens}$ for a source
at redshift $z_s=5$  (Turner, Ostriker and Gott 1984).  These results
are for  a  universe  with  a   mean  total density of $\Omega=1$ (the
Cosmological  Constant, $\Lambda=0$),  and for  a given  redshift the
ratio $\tau/\Omega_{lens}$ increases  as $\Omega$ decreases. The study
by  Schneider  (1993)   demonstrated  that  $\Omega_{lens}\sim0.5$  is
required (see  his figure 6b)  if the variability reported  by Hawkins
(1993) is  attributed  to   gravitational  lensing.   The  nanolensing
interpretation  of  quasar variability  is  also  consistent with  the
currently  favoured  combination   of  cosmological  model  parameters
($\Lambda\simeq0.7$, $\Omega\simeq0.3$)  provided that essentially all
of the matter  is in the form of  lenses, i.e $\Omega_{lens}\simeq0.3$
(Minty 2001).   The details of the cosmological  model are unimportant
here; for simplicity of calculation  we have therefore adopted a model
universe with $\Omega_{lens}=\Omega=1$.

The magnification pattern which is  introduced by the lenses takes the
form of  a caustic network, in  which the appearance of  any source is
influenced by a large number of lenses (Paczy\'nski 1986a); an example
of such a network is shown in  figure 2. In this figure we see a large
number of fold caustics -- a  fold is the lowest order catastrophe, at
which two  images merge --  joined by cusps (catastrophes  where three
images merge). In  the present paper we illustrate  the effects of GRB
nanolensing through  analytic and  numerical calculations of  a source
crossing a single fold caustic  (this section). We then (\S4) simulate
the  global  lightcurves expected  for  a  source  expanding across  a
caustic network such as that shown in figure 2.

In  the immediate  vicinity  of  a fold  caustic  the mapping  between
rectangular source coordinates $(x,y)$, and image coordinates $(X,Y)$,
is well  approximated by  a quadratic form  (see Schneider,  Ehlers \&
Falco 1992):
\begin{equation}
	x = {1\over2}aX^2,\qquad
        y   = {1\over2}Y,
	\label{eq:mapping}
\end{equation}
with  the $x$  axis oriented  perpendicular  to the  caustic, and  the
curvature,  $a$, being a  positive constant. Here the factor $1/2$ in
the $x\leftrightarrow X$ mapping is arbitrary (because $a$ remains
unspecified), but the factor of $1/2$ in the $y\leftrightarrow Y$ mapping
is required in order that the fold be convergence free -- a pure shear
lens mapping -- as all empty-beam lens mappings must be in General
Relativity. There are  two solutions
(images) to these equations for any source located at $x>0$, and it is
straightforward to determine their magnifications:
\begin{equation}
	\mu_\pm = \pm {{1}\over{2\sqrt{2ax}}}.
	\label{eq:magnification}
\end{equation}
In addition to this image pair,  there will typically be a large
number  of other  images  present, but  the  observed flux  variations
around the  time of caustic  crossing are completely dominated  by the
two very bright images which are  described by equations 12 and 13. If
a burst occurs at any location $x<0$, then neither of these two images
will be present  until the source expands so as  to touch the caustic,
at  which point the  observed flux  will rise  abruptly. On  the other
hand, if a  burst occurs at $x>0$ then both images  will be present at
all times.

There  are two  separate cases  which we  must consider,  depending on
which side  of the  caustic the  burst occurs. For  a point  source at
$x<0$ there are $2n+1$ images,  whereas for $x>0$ there are $2n+3$; we
therefore denote  these two regions  as the one-image  and three-image
regions, respectively, with the understanding that there are many more
images present, but  only the pair created/annihilated at  $X=0$ is of
interest here.  If the  source takes the  form of  a thin ring,  as is
approximately the case for the  model we have adopted (\S2; see figure
1),  then the  lensed  images appear  as  shown in  figure  3. In  our
subsequent development we  denote the $x$ coordinate of  the origin of
the burst  by $s$, so that the  sign of $s$ determines  whether one or
three images are initially present.

\subsection{Burst in single-image region ($s<0$)}
Because  the  source intensity  distribution  is  axisymmetric, it  is
useful to  consider the  properties of an  infinitesimally-thin ring,
radius $r$, of uniform  surface-brightness, under the mapping given by
equation 12. If  the burst occurs at $s<0$, then  images of the source
first appear when the source reaches a radius $r=|s|$, and for $r>|s|$
the total magnification of the ring (summed over both images) is given
by
\begin{equation}
\mu_1={{1}\over{\pi}}\sqrt{{u}\over{ar}}\;{\cal F}(\phi | u)
\end{equation}
where ${\cal F}$ denotes the Incomplete Elliptic Integral of the first
kind;     $\cos2\phi:=-s/r$,      and     $u\equiv2r/(r+s)$.      Here
``magnification'' means the  total area occupied by the  two images of
the  annulus,  divided by  the  area  of  the annulus  itself.  Simple
approximations    to     this    exact    result     are    available:
$\mu_1\simeq(1-3\epsilon/8)/2\sqrt{a|s|}$  for  $0<\epsilon\ll1$, where
$\epsilon\equiv    r/|s|    -    1$;    and    $\mu_1\rightarrow{\cal
K}(1/2)/\pi\sqrt{ar}$  for $r\rightarrow\infty$,  where ${\cal  K}$ is
the  Complete  Elliptic  Integral   of  the  first  kind,  and  ${\cal
K}(1/2)\simeq1.8541$.  The  exact form of the  magnification, given by
equation 14, is graphed in figure 4 for the case $s=-1/a$.

\subsection{Burst in three-image region ($s>0$)}
For a burst which occurs  in the three-image region ($s>0$), the total
magnification of an annulus is given by
\begin{equation}
\mu_3={{1}\over{\pi}}\sqrt{{u}\over{ar}}\; {\cal K}(u)
\end{equation}
for $r<s$. On the other hand for $r>s$ we have (as with eq. 14)
\begin{equation}
\mu_3={{1}\over{\pi}}\sqrt{{u}\over{ar}}\; {\cal F}(\phi | u).
\end{equation}
Simple  approximations  can be  found  for  these  expressions in  the
following    regimes:   $\mu_3\simeq1/\sqrt{2as}$,   for    $r\ll   s$;
$\mu_3\simeq\ln(32/|\epsilon|)/2\pi\sqrt{as}$,  where  $|\epsilon|\ll1$
(and $\epsilon=  r/s-1$); and $\mu_3\simeq{\cal K}(1/2)/\pi\sqrt{ar}$
for $r\rightarrow\infty$. The exact results are shown in figure 4, for
the  case  where $s=1/a$,  alongside  the corresponding  magnification
curve for a burst occurring in the single image region.

\subsection{The median lens}
For some  purposes (e.g. estimating  a characteristic length  scale in
the  source  plane)  it  is  important  to  know  whereabouts  on  the
line-of-sight the typical lens is  located. This need is most sensibly
addressed by  determining the redshift  at which the optical  depth to
gravitational lensing is  one half of the total  optical depth for the
source under  consideration. Using equation 2.13b  of Turner, Ostriker
and Gott (1984) this can  be quite straightforwardly achieved, and the
result  is that  for a  source at  redshift $z_s$  the median  lens is
located at redshift $\langle z\rangle$:
\begin{equation}
\langle1+z\rangle=\sqrt{1+z_s}.
\end{equation}
Thus  for a source  at redshift  $z_s=3$ the  median lens  redshift is
unity,  and at  low redshift  ($z_s\ll1$) the  median lens  is located
half-way to the source.

One  application   of  the  foregoing  is   the  relationship  between
transverse  dimensions  measured  in  the  source  plane  and  in  the
observer's  plane.   Appropriate  angular-diameter distances  for  our
circumstance are evidently ``empty-beam'', distances (for $\Omega=1$),
and  these  are  given  in  table  1  of  Turner,  Ostriker  and  Gott
(1984).   Denoting  observer-lens  and   lens-source  angular-diameter
distances  by  $D_d$  and  $D_{ds}$, respectively,  we  introduce  the
``lever arm'', $\varpi\equiv D_{ds}/D_d$.  It is this ratio which, for
any given  lens, determines the relationship of  a transverse distance
in the  source plane to  the corresponding distance in  the observer's
plane.  It  is straightforward  to  show  that  the median  lever  arm
(i.e. the lever arm of the median lens) is
\begin{equation}
\langle\varpi\rangle=(1+z_s)^{-3/4},
\end{equation}
and we will make use of this result in \S5.

\subsection{Variability time-scale}
In  combination  with our  kinematic  model  of  the source  structure
(\S2.1), the  statistical properties of the  caustic network determine
the predicted variability  time-scale of GRBs as a  function of source
redshift and lens  mass. At each occasion when the  limb of the source
crosses a caustic,  a peak is introduced into  the light-curve, so the
characteristic time-scale between peaks is  just the time taken by the
source to expand  in area by $1/\Sigma$, where  $\Sigma$ is the number
of caustics  per unit  area in  the source plane.  In the  low optical
depth regime  there are  four folds in  each projected
Einstein  Ring.  At  high  optical  depth,  where  a  caustic  network
develops, the caustic density is higher  by a factor equal to the mean
magnification,  $(1-\tau)^{-2}$,  and  we  therefore  estimate  the
caustic  density by  $\Sigma\sim4\tau/(1-\tau)^2$ per  unit Einstein
Ring area  in the source plane. The  circumstance $\tau=1$ corresponds
to  $z_s\simeq10.7$ in  our model  universe, and  the  caustic density
becomes very large,  and variability time-scale correspondingly short,
for sources around this redshift.

Using our self-similar ($m=12$) source model, we can now estimate the
expected variability time-scale, $t_{var}$ via
\begin{equation}
t_{var}^{-1} \sim \Sigma{{\rm d\ }\over{{\rm d} t}}\;\pi r_{max}^2.
\end{equation}
It  is  worth  noting  that  $r_{max}^2\propto  t^{(m+2)/(m+1)}$  (see
eq.  3), so  that the  apparent area  of the  source  increases nearly
linearly  with  time,  and  the  variability time-scale  is  thus  not
expected  to evolve  significantly during  a burst.   The  results are
shown in figure 5 as a function of lens mass, for two different values
of  the source  redshift ($z_s=1,5$),  and an  initial  Lorentz factor
$\Gamma_0=10^3$.  Considering  that  bursts  may last  up  to  several
hundred seconds,  and that the available temporal  resolution is below
100~ms in existing data, almost  the entire temporal range in figure 5
is  open  to  study.  We  thus  recognise  that  {\it  a  cosmological
population  of planetary-mass lenses  should introduce  variability to
gamma-ray  burst  profiles,\/}  suggesting  a  powerful  test  of  the
nanolensing interpretation of quasar variability.

The model  does not  make a precise  prediction for  the time-interval
between caustic  crossings, because the  appropriate lens mass  is not
tightly constrained (nor indeed are the parameters of the source, such
as  $\Gamma_0$).   Hawkins  (1993)  initially  gave   an  estimate  of
$10^{-3}\msun$,  but  Schneider's   (1993)  analysis  clearly  favours
$10^{-4}\msun$  or even  smaller values.  Using  three-dimensional ray
shooting simulations, Minty (2001)  finds that these data suggest lens
masses of  $10^{-5}-10^{-4}\msun$. Referring to figure 5,  if we adopt
$\Gamma_0=10^3$  and lens  masses  of order  $10^{-5}\msun$, then  the
variability time-scale is expected to be $\sim60$~seconds for a source
at  redshift  $z_s\simeq1$,  and   $\sim4$~seconds  for  a  source  at
$z_s\simeq5$.

\section{Light curves}
We  have  taken  two  distinct  approaches  to  the  study  of  lensed
light-curves: detailed calculation of the behaviour around the time of
a fold-caustic-crossing  event, exhibiting the type  of profiles which
are expected  for individual ``pulses''  (cf. Norris et al  1996); and
simulation   of  light-curves   for  GRBs   seen  through   a  caustic
network. For any given burst  the latter reflects the random structure
of the  caustic network in  the vicinity of  the source. We  note here
that the propagation times for  the various images in our model differ
only by amounts $\la1$~ns, and this is negligible for our purposes.

\subsection{Individual ``pulses''}
Making use  of the source intensity  profile computed in  \S2, and the
magnification  curves derived  in the  previous section,  it is  now a
straightforward exercise to compute the flux history, $F(t)$, from
\begin{equation}
F(t)\propto\int\!\!{\rm d}r\,2\pi\,r\,I(r,t)\,\mu,
\end{equation}
where the total magnification, $\mu$, can be written as the sum of the
``unlensed'' images  (actually, a large number of  lensed images whose
total  flux is  roughly constant  over the  time-scale of  the caustic
crossing), plus  the bright  pair of images  associated with  the fold
caustic under consideration:  $\mu=1+\mu_{1,3}$.  (The choice of unity
for the constant term in this relation is somewhat arbitrary.) As with
equation 11, this formula can  easily be rewritten as an integral over
the variable  $q$. After  substituting for $I(r,t)$  (equations 7--10)
and $\mu_{1,3}$  (equations 14--16) it is  straightforward to evaluate
equation 20 numerically;  to do this we used  the Mathematica software
package. The  results are  shown in figure  6, for bursts  occurring in
both  one- and  three-image regions.  For these  calculations  we have
assumed $|s|=1/a$,  and we have taken the  fractional radial thickness
of  the  emitting  shell   to  be  $\Delta=4\times10^{-3}$.  For  each
light-curve,  the limb  of the  source  first touches  the caustic  at
$t=1.5$~seconds (an arbitrary choice).


\begin{figure*}
\plotone{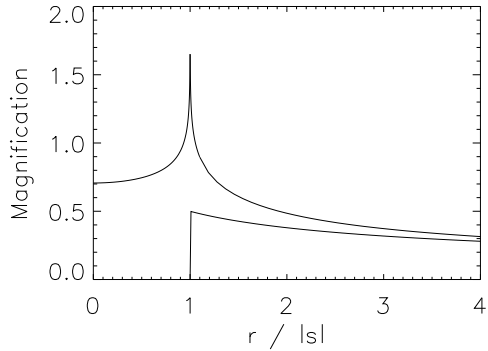}
\caption{The   magnification  of  an   infinitesimally-thin,  annular
source, of  radius $r$, centred on  $x=s$, under the  mapping given in
equation  12;  this  mapping  describes a pure shear fold  caustic  at
$x=0$. Here  we have  assumed that the  curvature of the  lens mapping
near  the fold  is $a:=1/|s|$.  The upper  curve corresponds  to $s>0$
(burst occurs in three-image  region), and the lower curve corresponds
to $s<0$ (burst in single-image region).
\label{fig:magnification}}
\end{figure*}

\begin{figure*}
\plotone{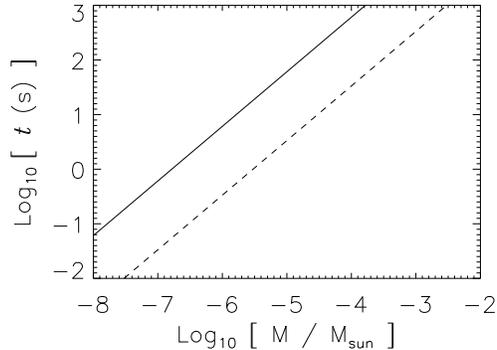}
\caption{The  nanolensing variability time-scale,  in a  Universe with
$\Omega_{lens}=\Omega=1$, as  a function of lens mass.   The source is
assumed to be in self-similar, fully-radiative ($m=12$) evolution with
initial    Lorentz    Factor    $\Gamma_0=10^3$;   other    parameters
characterising  the  expansion  have  been fixed  at  $E_{52}=1$,  and
$n=1$. The  solid line  corresponds to a  source at  redshift $z_s=1$,
while the dashed line is for $z_s=5$.}
\end{figure*}

\begin{figure*}
\plotone{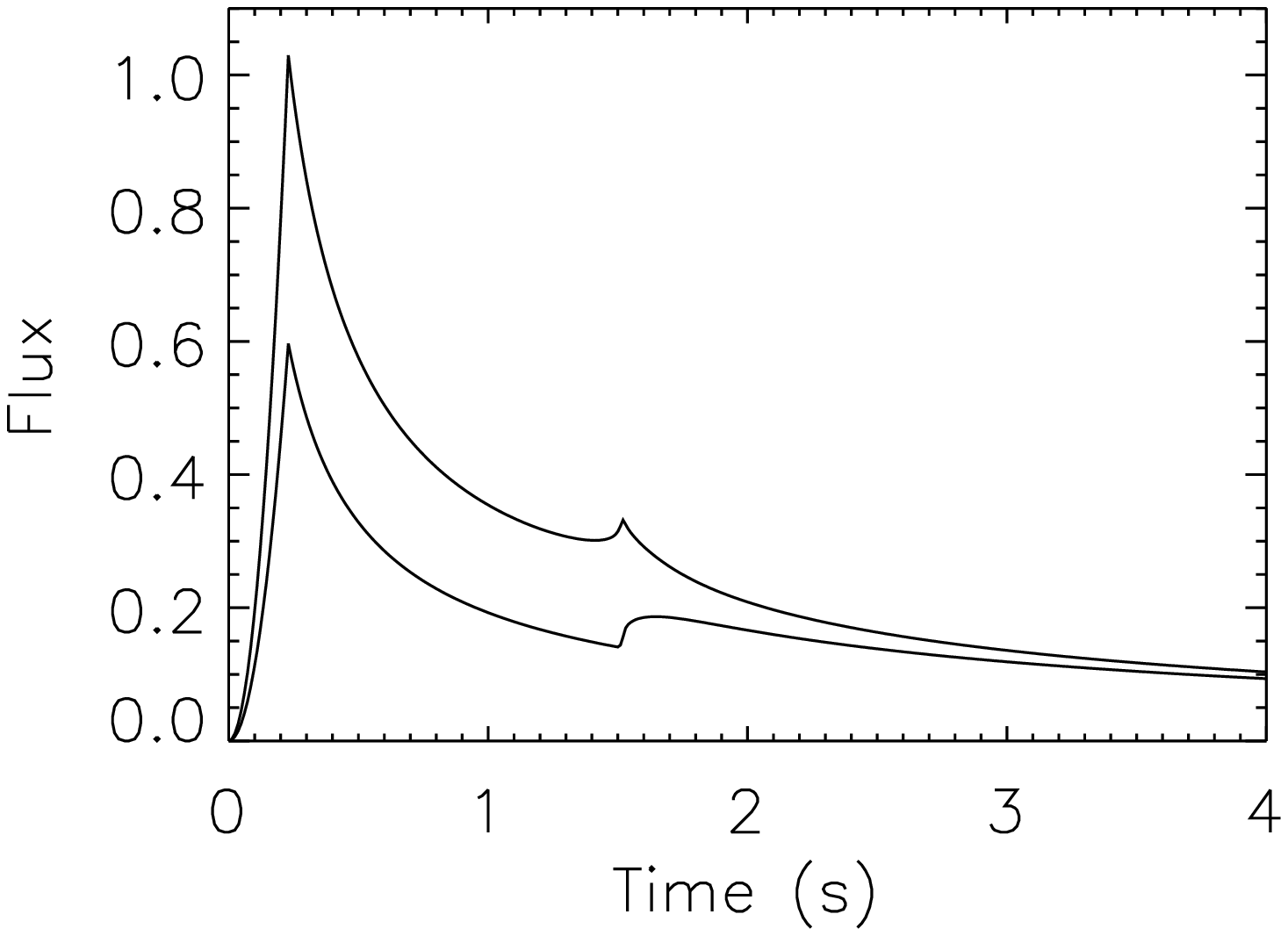}
\caption{Bolometric  $\gamma$-ray  light-curves  for  a  relativistic,
fully-radiative  ($m=12$),   self-similar  blast-wave  seen  expanding
across a fold  caustic. The thickness of the  radiating shell is taken
to  be $\Delta=4\times10^{-3}$,  and the  centre of  expansion  of the
burst is  assumed to be a distance  $1/a$ from the fold,  where $a$ is
the curvature of  the lens mapping.  The upper  curve corresponds to a
burst in the three-image region, while the lower curve is for an event
in the single-image region; caustic crossing occurs at $t\simeq1.5$~s.}
\end{figure*}


Some aspects of figure 6  require immediate comment. The difference in
overall normalisation of the two curves at early times simply reflects
the ratio  of total magnifications  at the onset  of the burst  -- see
figure 4 -- roughly  $(1+1/\sqrt{2})/(1+0)\simeq1.7$. This property is a
direct  result of  the simple  lens model  we have  adopted  (i.e. the
quadratic relationship  between source and image  coordinates), and it
should be remembered  that this approximation is accurate  only in the
immediate vicinity  of the caustic.  Consequently it is  the behaviour
seen in the light-curves around  $t=1.5$~s which is of prime interest,
rather than the global properties. The essential features to note are,
therefore,  that: (i)  the presence  of  a fold  caustic introduces  a
significant peak into the light-curve,  close to the time at which
the limb of  the source first touches the caustic,  and (ii) this peak
can be either cuspy,  and approximately time-symmetric, or rounded and
asymmetric.  These  features could,  in  fact,  have been  anticipated
simply  by examining  the magnification  profiles shown  in  figure 4,
bearing in  mind that  the source structure  is basically just  a thin
ring (figure 1).

\subsection{Simulated light-curves}
Although  the  circumstance we  are  considering  is  one in  which  a
population  of gravitational  lenses is  distributed along  the entire
line-of-sight,   we   have  made   use   of   simulations  which   are
two-dimensional, with  all of  the lenses located  in a  single plane.
This is  partly because three-dimensional  ray-tracing simulations are
much  slower  than  their  two-dimensional  counterparts,  but  mainly
because  a  well-tested  two-dimensional  code  was  available  to  us
(Wambsganss, Paczy\'nski \& Katz 1990). Two extensions of this code were
required for  our purposes: convolution of the  magnification map with
our model  source intensity profile (\S2); and  a progressive ``zoom''
which recomputes the magnification map  on larger scales as the source
expands.   The  latter feature  was  implemented  with an  incremental
``zoom'' factor of 1.2, and  our computational grid was 2048 pixels on
a side, so that our source radius  was in the range 800 to 1000
pixels.  Despite  this our  resolution is only  just adequate  for the
task, as the reader can easily verify by examining the radial width of
the  source intensity  distribution (figure  1).  An  estimate  of the
Full-Width at Half-Maximum (FWHM) of  the source intensity peak can be
obtained from
\begin{equation}
{\rm   FWHM}\;\simeq\left(   {{m+2}\over{2m+3}}  \right)^{(m+2)/(m+1)}
 {{2\sqrt{3\Delta}}\over{2m+3}},
\end{equation}
and for  $m=12$, $\Delta=4\times10^{-3}$ this  yields (coincidentally)
${\rm FWHM}\simeq4\times10^{-3}$.  With a  source radius of 800 pixels
the intensity profile is thus 1.5 times over-sampled.


\begin{figure*}
\plotone{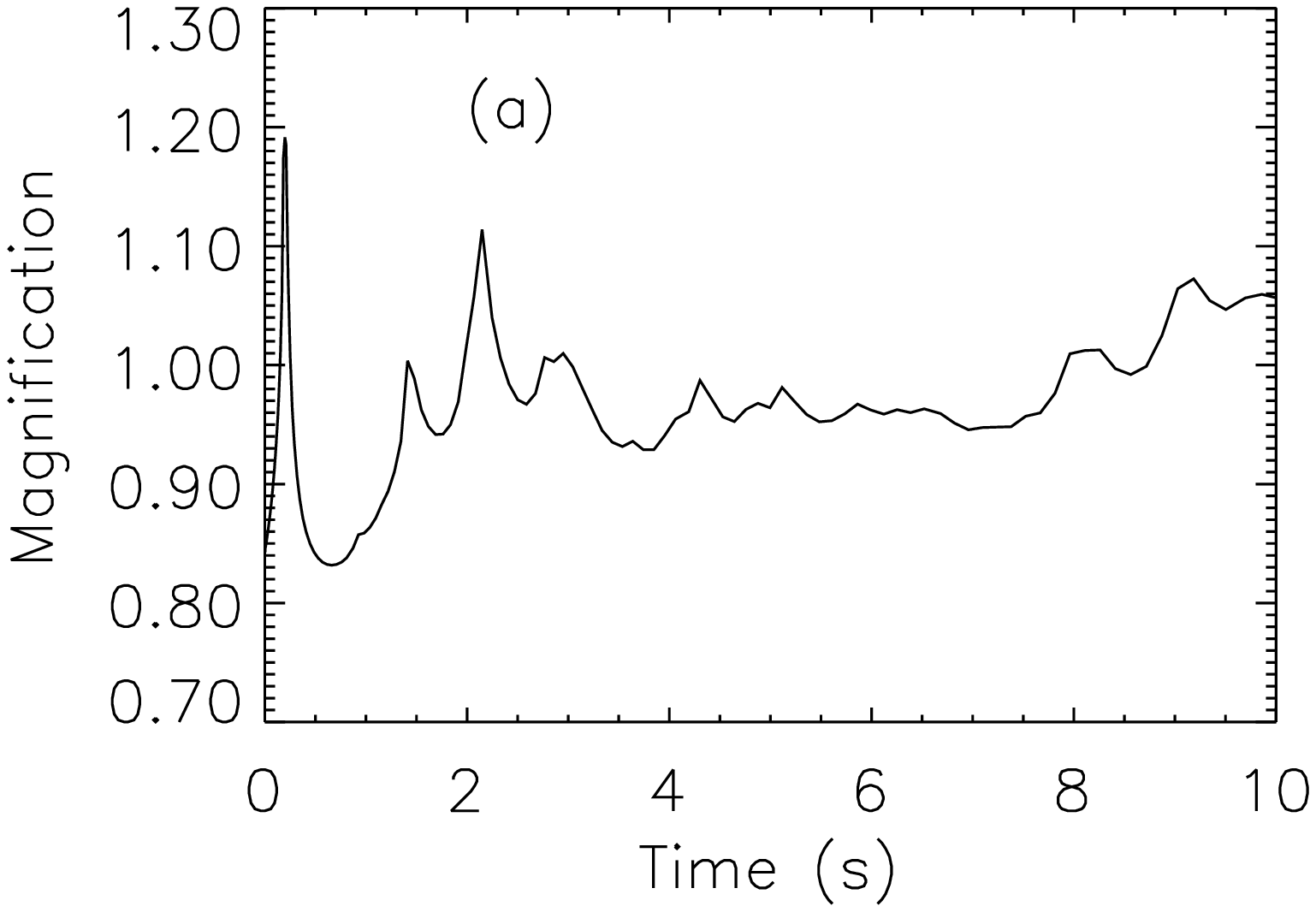} \plotone{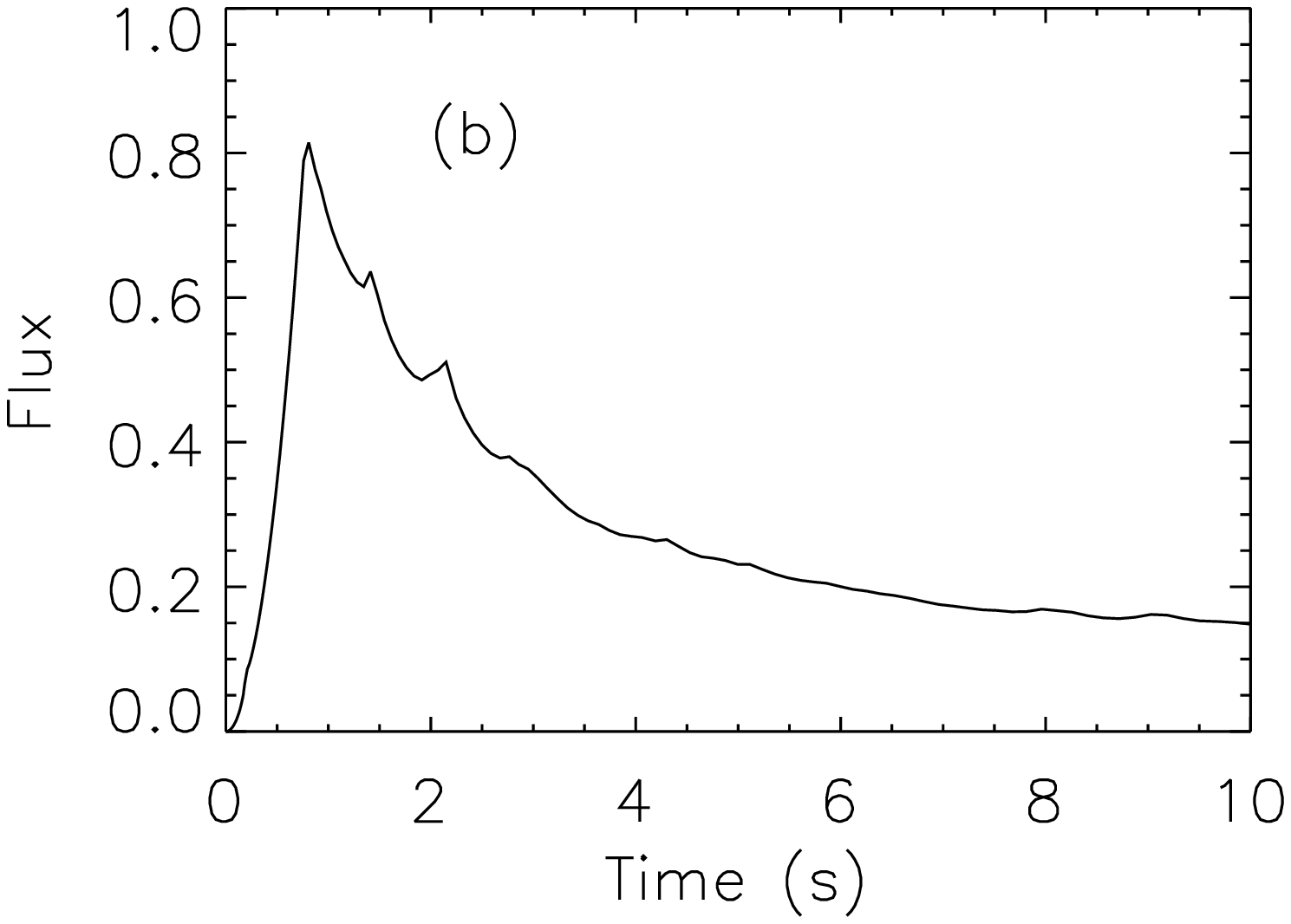}
\caption{Simulations of the mean magnification (panel a) and resulting
light-curve (panel  b) for a GRB  occurring at redshift  $z_s=1$, in a
universe full  of nanolenses ($\Omega=\Omega_{lens}=1$);  each lens is
assumed  to have  a  mass of  $10^{-8}\;{\rm  M_\odot}$. The  apparent
time-scale of the transition  between coasting and radiative phases is
given  by  $(1+z_s)\,R_0/c\Gamma_0^2$,   and  for  our  adopted  burst
parameters     ($E_{52}=n=\Gamma_3=1$)     this     corresponds     to
$0.39\,(1+z_s)$~s. This time-scale defines the location of the peak of
the burst in panel (b). Note that the magnification is in units of the
theoretical  average, $\left<\mu_{th}\right> =  (1-\tau)^{-2}$, which,
for this case, is 1.26.}
\end{figure*}

\begin{figure*}
\plotone{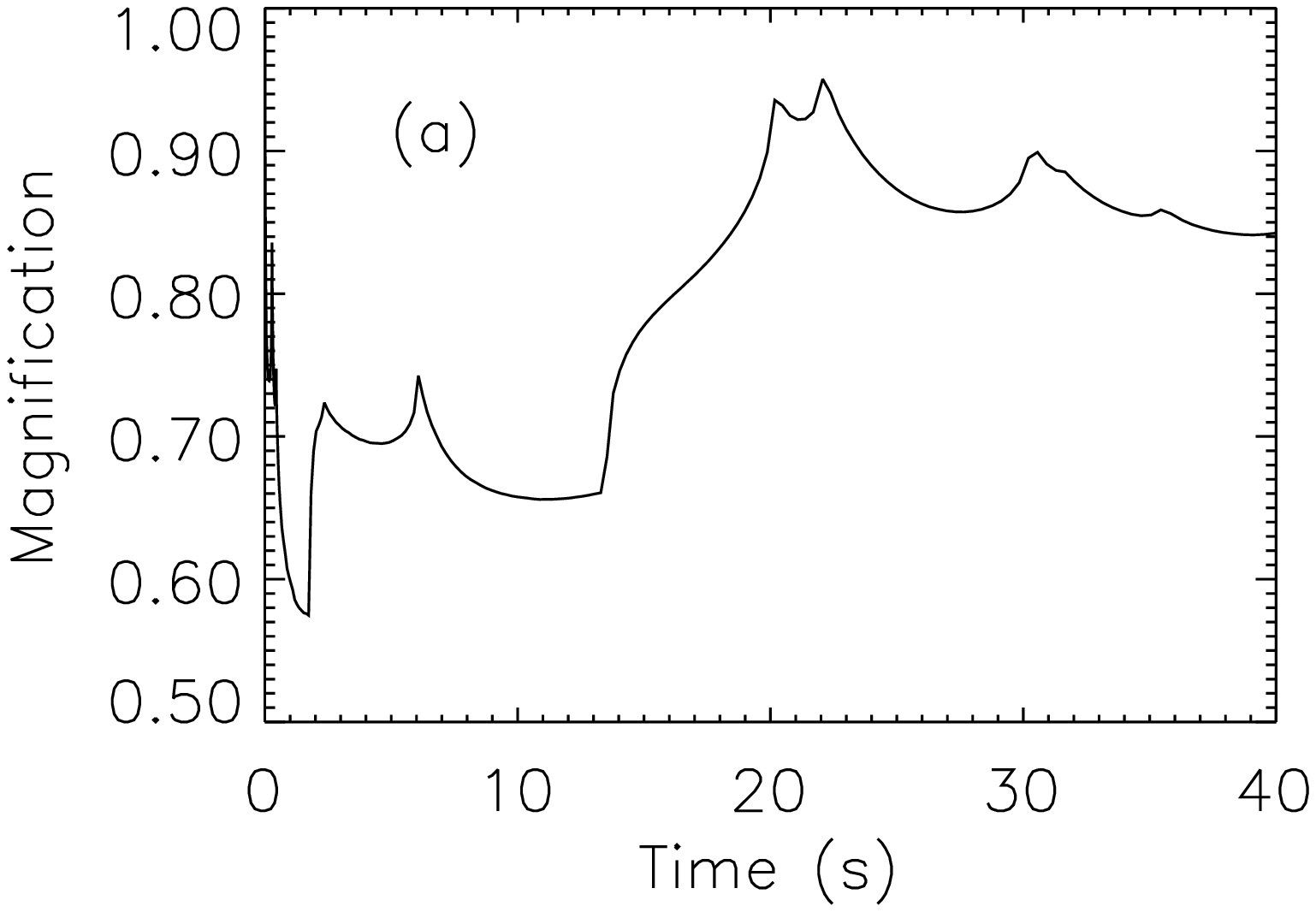}\plotone{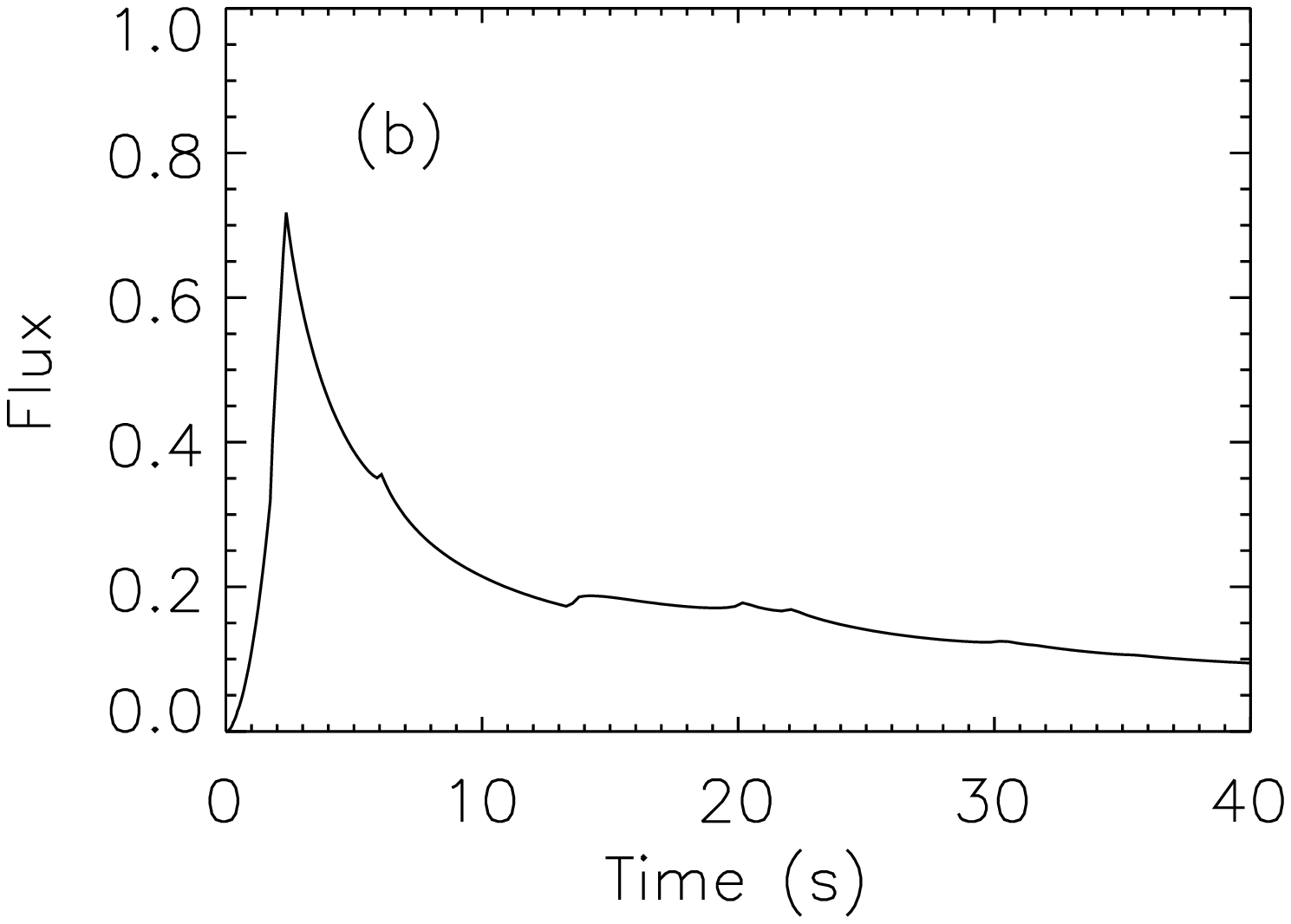}
\caption{As  figure   7,  but  for   a  source  redshift   of  $z_s=5$
(corresponding to  the burst  peak at $t=2.3$~s),  and lenses  of mass
$10^{-5}\;{\rm M_\odot}$. Again, the  magnification is in units of the
theoretical average, $\left<\mu_{th}\right>=6.9$.}
\end{figure*}


To illustrate the type of lightcurves which might be expected, we have
undertaken simulations  for sources  at redshift $z_s=1$  and $z_s=5$,
corresponding  to  optical  depths  to  gravitational  nanolensing  of
$\tau=0.11$  and  $\tau=0.62$,  respectively.  The  results  of  these
simulations  are  shown  in  figures  7  and 8,  for  lenses  of  mass
$10^{-8},10^{-5}\;{\rm  M_\odot}$,  respectively.  Because the  source
evolves in a self-similar (scale-free)  manner, it is not necessary to
recompute magnification curves when the adopted lens mass is  changed:
different lens masses simply require a rescaling  of the time  axis in
panel (a) of these figures, thus changing the characteristic time-scale
of any variability, but not its amplitude. The choice of lens mass was
made, for each of the two redshifts, so as to maximise the variability
within each of the synthetic bursts.

All of our simulations follow the same basic pattern: starting from an
initial  magnification which  is below  unity, the  mean magnification
(i.e.  average magnification  over  the whole  source) exhibits  rapid
variability,  the   amplitude  of   which  decreases  as   the  source
expands. These  features are readily understood.  Initially the source
is likely  to be demagnified, because  most of the area  of the source
plane    is   occupied   by    regions   with    magnification   below
unity.  Subsequently,  as  the  source expands  across  the  caustics,
variability  is seen.  However,  accompanying the  increase in  source
radius, is an increase in the thickness of the high intensity ring and
the modulation of the mean magnification of the source is consequently
reduced. Simultaneously the time-averaged magnification trends towards
its large-scale average value as the source covers an increasing area.

There are two important points  which are evident on examining figures
7  and 8. First,  the time-scale  of the  variations in  the simulated
light-curves is in  broad agreement with the rough  estimates given in
\S3.4 (see  figure 5). Secondly, the  depth of modulation  seen in our
simulated  magnification curves  is  quite small,  being typically  of
order  10\%, and  in the  light-curves themselves,  where  the secular
evolution  is strong,  it  is not  always  apparent that  the flux  is
varying on short time-scales.

\subsection{Comparison with observations}
Many high-quality  gamma-ray burst  light-curves were acquired  by the
BATSE experiment on the Compton Gamma-Ray Observatory (e.g. Fishman et
al  1994), and  the substructure  within these  light-curves  has been
analysed by  Norris et  al (1996:  N96), in the  case of  long, bright
bursts.   The principal  conclusion of  this analysis  was  that these
bursts can  be regarded  as being  made up of  a number  of ``pulses''
having a  typical separation of order  one second (but  which are {\it
not\/} periodic).  N96 found  that the typical  pulse shape  is cuspy,
similar to the peak near $t\simeq1.5$~s in the top curve in figure 6.

Comparing these  findings to the  results of \S\S3,4, we  can identify
three key points.  First, the  observed time-scale of order one second
between   ``pulses''  is  comparable   to  the   predicted  time-scale
(\S\S3.4,4),   {\it   but  only   if   the   lenses   have  low   mass
($\sim10^{-8}\;{\rm M_\odot}$), or the sources are at high redshift.\/}
Secondly, the  typical ``pulse'' shape  observed by N96 is  similar to
the   $3\rightarrow1$,  i.e.   image-annihilating,   caustic  crossing
discussed in \S3.2, and shown in  figures 4 and 6. However, the theory
presented  earlier  in this  paper  gave  no  reason to  suppose  that
$3\rightarrow1$  crossings should  be  preferred over  $1\rightarrow3$
crossings, and  in fact one does  expect both to be  present.
Both {\it are\/}
present in  the simulated light-curves shown  in figures 7  and 8. One
could appeal  to an  observational bias which  favours $3\rightarrow1$
caustic crossings over their $1\rightarrow3$ counterparts, because the
high and  cuspy peaks of  the former render  them more visible  in the
light-curves (see,  particularly, figure  4).  However, this  does not
seem  to us  to be  a strong  argument. Finally,  although  the single
fold-crossing events modelled in  \S3 are obvious in the light-curves,
it is plain  that {\it the simulations in \S4  do not yield sufficient
modulation to be  able to explain the profound  variations seen in the
data.\/}

On the last  of the above points we note that  there is no discrepancy
between  the calculations  undertaken in  \S3 and  the  simulations of
\S4. The difference  between the computed modulation depth  in the two
cases  simply reflects  the  fact  that the  caustics  present in  the
simulations  exhibit  more curvature  in  the  lens  mapping than  was
assumed in \S3.

\section{Parallax}
A key aspect of the lensing model we are investigating here is that of
parallax. The importance of  parallax phenomena depends on the spatial
separation of the detectors which are employed, in comparison with the
spatial scale on  which the magnification pattern changes.  In the low
optical depth regime, the latter scale is given by
\begin{equation}
\left[4{{GM}\over{c^2}}{{D_sD_d}\over{D_{ds}}}\right]^{1/2}\simeq
60\sqrt{M_{-4}}\quad{\rm AU},
\end{equation}
where $D_s$  is the angular-diameter  distance of the source  from the
observer,  and to  arrive  at  the numerical  result  we have  assumed
$z_s=3$ and considered the median lens.  Hence for detectors separated
by a  few Astronomical  Units, parallax might  be expected to  show up
only for lenses of  mass $M\la10^{-7}\msun$ (Nemiroff and Gould 1995).
However, in the model we  are considering the expanding source crosses
caustics, and  in the vicinity  of caustics the  magnification pattern
varies  significantly on  spatial  scales   much  smaller  than  given 
by  equation 22.
Indeed  at  caustic  crossings  the transverse  flux  gradient  is
limited only by  the source structure, and parallax  may be observable
even over  very modest  baselines. Specifically, if  the width  of the
high intensity ring  of the source is taken  to be $4\times10^{-3}$ of
the apparent radius (see \S4.2),  and the latter is approximately 6~AU
at  $t=10$~seconds  ($E_{52}=n=\Gamma_3=1$),   then  parallax  may  be
evident over transverse length scales as small as $4\times10^{11}$~cm.
In  this circumstance the  main instrumental  requirement is  for good
temporal resolution in the  detectors, and high signal-to-noise ratio,
so that  the times at which  caustic crossings occur  can be precisely
determined (Grieger, Kayser and  Refsdal 1986; Hardy and Walker 1996).
Gamma-ray bursts  are typically  recorded with temporal  resolution of
tens  of milliseconds, or  better, often  at high  signal-to-noise, so
that   the   available    timing   precision   is   generally   rather
good.  Moreover,  we  wish  to  emphasise  that,  provided  there  are
significant increases in flux  associated with a caustic crossing, the
ability to precisely time such  an event is essentially independent of
lens mass and  this parameter does not influence  the detectability of
the two parallax phenomena we shall describe.

\subsection{Displaced burst locations}
Suppose we  observe a burst with  a pair of  identical detectors whose
spatial separation is small in  comparison with the scale on which the
lens magnification  pattern changes.  The caustic network  will appear
very  similar as seen  from the  two different  locations, but  with a
slight shift on the sky  in a direction parallel to the sky-projection
of the vector separation  between the detectors, $\vec b_{12}=\vec x_1
- \vec x_2$,  and by an amount proportional  to $b_{12}$. Consequently
the  two  light-curves are  almost  identical  in  structure, but  the
caustic-crossings are seen to occur at slightly different times at the
two  detectors. Thus  if burst  time-of-arrival is  used  to determine
which direction the wavefront has arrived from, then the present model
predicts that the inferred location  of the source is shifted from the
true location (cf. McBreen and Metcalfe 1988).

In the  absence of  any lensing phenomena,  under the assumption  of a
source at infinity (plane wavefront), a measured difference in arrival
time,  $\Delta t_{12}=t_2-t_1$,  between detectors  tells us  that the
burst came from a direction defined  by the unit vector $\hat n$, such
that  $\hat n\cdot\vec  b_{12}=c\,\Delta t_{12}$.  Thus the  source is
constrained    to    lie   on    a    circle    of   angular    radius
$\alpha:=\cos^{-1}(c\,\Delta t_{12}/b_{12})$, centred on the direction
$\hat b_{12}$. If a timing offset, $\delta t_{12}$, is introduced then
the  corresponding   angular  offset,  $\delta\alpha$,   is  given  by
$\sin\alpha\,\delta\alpha=-c\,\delta    t_{12}/b_{12}$.   The   timing
difference which is expected as a result of parallax, in our model, is
roughly  $\delta t_{12}\la\varpi\sin\alpha  (b_{12}/\beta_{ap}c)$, for
any  given caustic  crossing. (The  precise value  of  $\delta t_{12}$
depends, of  course, on the  orientation of $\vec b_{12}$  relative to
the orientation of the caustic.) Thus, for a burst which exhibits only
a single caustic crossing event, the expected angular offset is
\begin{equation}
|\delta\alpha|\la{{\langle\varpi\rangle}\over{\beta_{ap}}}.
\end{equation}
Some caution should  be exercised in applying this  result, because in
our case, where a network of  caustics is present, the folds cannot be
readily associated with individual  lenses, and because the lenses are
distributed along  the whole  line of  sight it is  not clear  how the
effective value of $\varpi$ should be estimated. We shall nevertheless
employ equation  23 as  if the values  of $\varpi$ were  in one-to-one
correspondence with  lenses, and use  the median value of  $\varpi$ as
given in  eq. 18.  For  a source at  redshift $z_s=3$, at  an observed
time $t\sim10$~s after the start of  the burst, equation 6 leads us to
expect  an  apparent  expansion  speed of  $\beta_{ap}\simeq320$,  and
projecting  this   into  the   observer's  plane  yields   the  result
$|\delta\alpha|\la4$~arcmin. This estimate is  valid for a burst which
exhibits only  a single  caustic crossing; if  the burst  has temporal
substructure  with $N_p$  peaks  in the  light-curve,  then we  expect
$|\delta\alpha|$ to be  smaller by a factor of  order $\sqrt{N_p}$. We
now turn to the question  of whether such systematic errors are either
evident in existing data, or are significantly constrained by them.

Comparing burst  arrival times between three  spacecraft localises any
source  to one of  two possible  intersections between  three separate
loci  (one  locus derived  from  each  spacecraft  pair). Because  the
observable quantity is the burst arrival time, and each locus reflects
the difference between a pair of arrival times, only two of these loci
provide independent information on  the burst location; in other words
the  location is  not determined  with  any redundancy  if only  three
spacecraft are employed. If four or more interplanetary spacecraft are
available  for  burst  triangulation,  then redundant  positioning  is
possible, and  systematic errors can therefore be  revealed using only
the triangulation data. Many  bursts were redundantly positioned using
the  first  generation  of  interplanetary gamma-ray  burst  detectors
(Atteia  et al  1987), and  in two  of these  cases there  were highly
significant  discrepancies  (ten   standard  deviations)  between  the
localisations.   However,   in  at  least  one  of   these  two  cases
(GRB~790329)  the cause of  the discrepancy  appears to  be understood
(Laros et al 1985), and  we adopt the conservative assumption that the
other discrepant  burst (GRB~790116)  is also attributable  to effects
other than the parallax  phenomenon under discussion here. The typical
level  of  agreement between  redundantly  determined burst  locations
excludes errors  at the level $|\delta\alpha|\ga1^\circ$  (Laros et al
1985), but this  result does not strongly constrain  the model we have
presented  because   offsets  as large as  this are expected  only  in
exceptional circumstances. Specifically,  offsets larger than a degree
can only occur if the apparent expansion speed is $\beta_{ap}<60$ {\it
and\/} the source  is at low redshift ($z_s<1$)  {\it and\/} a caustic
crossing event occurs.

There are no four-spacecraft  triangulations available from the modern
(BATSE and  post-BATSE) era  of gamma-ray burst  studies, and  we have
therefore  searched   the  literature  for  bursts   which  have  been
accurately located both by triangulation and by an independent method.
The  best sample  of this  type  that we  found is  comprised of  nine
gamma-ray bursts which were observed  by BATSE, Ulysses and either the
Pioneer Venus Orbiter  (Laros et al 1998) or  the Mars Observer (Laros
et  al  1997)  --  for  which  triangulated  positions  are  therefore
available  --  all  of  which  were independently  positioned  by  the
rotation modulation collimator of  the WATCH experiment (Sazonov et al
1998). The  smallest error associated with  the triangulated positions
($1\sigma$   statistical  plus   systematic)  for   these   bursts  is
20~arcseconds,    while    the    median    error    is    close    to
3~arcminutes.  Moreover  the  WATCH   error  circles  all  have  radii
exceeding      14~arcminutes      ($1\sigma$     statistical      plus
systematic).  Considering  that  we  expect  deviations  of  order  an
arcminute the WATCH  error circle, in particular, is  so large that at
first sight  it seems hopeless to  attempt to constrain  the model via
astrometry.  The situation is  not quite  as bad  as it  first appears
because  pairs   of  Inter-Planetary  Network   (IPN)  loci  sometimes
intersect at  a very acute  angle, in which  case the location  of the
intersection point is  very sensitive to any errors  in $\alpha$, thus
allowing us to  gauge the magnitude of the IPN  errors. There are four
examples  of  this  type   of  configuration  amongst  our  sample  of
bursts.  However, none of  the implied  errors is  highly significant,
when compared to the estimated measurement error, nor do these results
provide  any powerful constraints  on the  theoretical model.  We have
therefore relegated  the details of  our astrometric analysis  of this
sample to an appendix, and in the present section we confine ourselves
to a brief discussion of the outcome.

For  the four  bursts for  which we  could gauge  the actual  error in
$\alpha$, the mean error value is only $1.2$ times the estimated error
($1\,\sigma$ statistical-plus-systematic), and is thus consistent with
expectations. It is therefore appropriate  to quote the results of our
analysis in  the form  of upper limits.  To do  this we take  an upper
limit of  $3\,\sigma$ in each of  the four cases; in  order to compare
with our  theoretical prediction (equation  23) we then convert  to an
upper limit on the typical parallax error on each caustic crossing, by
multiplying the  $3\,\sigma_\alpha$ limit by  $\sqrt{N_p}$. The values
of  $N_p$ for  BATSE451 and  BATSE1698 were  taken from  Norris  et al
(1996): $N_p=4$ and $N_p=7$, respectively. For BATSE2387 and BATSE907,
examination of  the archival light-curves (http://cossc.gsfc.nasa.gov)
reveals  that $N_p=1$, and  $N_p\simeq5$, respectively.  The resulting
limits on parallax  for a single caustic crossing  are thus deduced to
be  $|\delta\alpha|< 2.7,\,37,\,2.6,\,37$~arcmin, from  BATSE451, 907,
1698 and 2387,  respectively. Even the lowest of  these limits is only
comparable  to  the  predicted  value  ($\la4$~arcmin  for  a  caustic
crossing occurring at $t=10$~s, and  a source at redshift $z_s=3$), and
so  we conclude  that  existing gamma-ray  burst  astrometry does  not
significantly constrain the model we have presented.


\begin{figure*}
\plotone{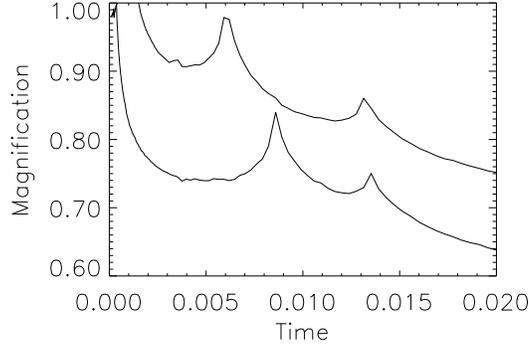}
\caption{The effect of parallax on the nanolensing magnification curve
of     an      expanding,     self-similar     ($m=12$)     blast-wave
($\Delta=4\times10^{-3}$).  Time is shown  here in  units of  the time
taken for  the source  to expand  to an apparent  radius equal  to the
Einstein  Ring  radius.  The  two  curves  differ  in respect  of  the
apparent source  location, which is  shifted by 1.4\% of  the Einstein
Ring  radius  (and  the upper  curve  has  been  displaced by  0.1  in
magnification,  for  clarity).  This  is equivalent  to  shifting  the
observer's   location   by  an   amount   1~AU,   transverse  to   the
line-of-sight, if  the lens mass  is $10^{-4}\;{\rm M_\odot}$  and the
source redshift is $z_s=5$.}
\end{figure*}

\begin{figure*}
\plotone{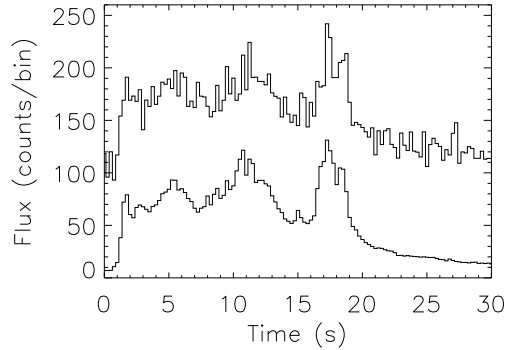}
\caption{The  Ulysses data  for  GRB920622 (BATSE1663;  Greiner et  al
1995), together  with the  light-curve derived from  the sum  of count
rates from BATSE channels 1 and 2. The BATSE data have been background
subtracted, then aligned and scaled appropriately by cross-correlating
with the Ulysses data, as described in \S5.2. The BATSE data have been
rebinned  to match  the binning  of the  Ulysses data,  and  have been
offset   by  $-100$~counts/bin  for   clarity  of   presentation.  The
signal-to-noise ratio of  the BATSE data is very high,  and all of the
features visible in the BATSE  light-curve are real. The flux scale is
appropriate to the Ulysses data.}
\end{figure*}

\begin{figure*}
\plotone{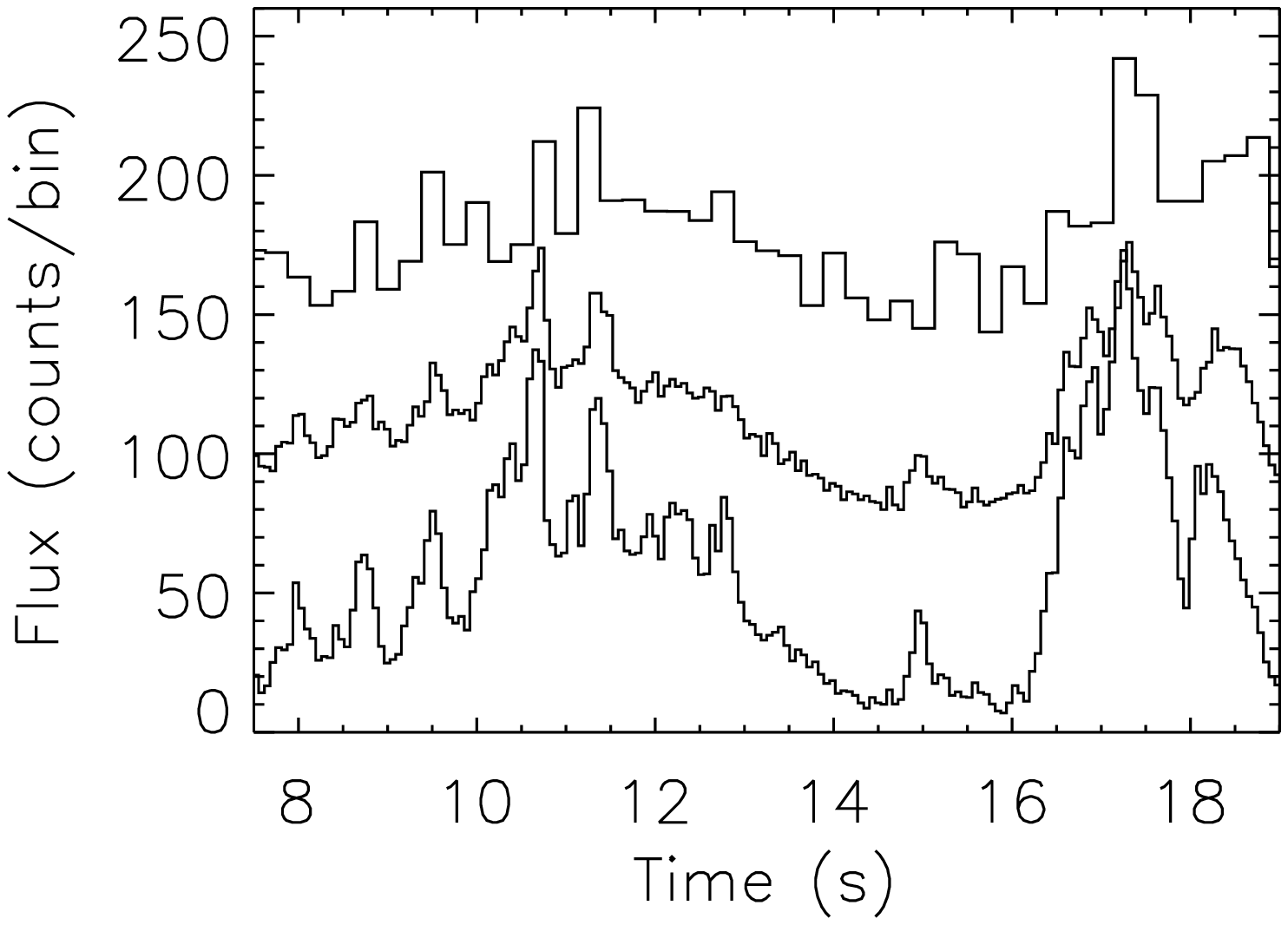}
\caption{A portion  of the two  light-curves shown in figure  10. Here
the BATSE high energy channels are shown (bottom curve) in addition to
the low energy data (middle  curve), and the Ulysses data (top curve);
both  BATSE  light-curves  are  shown  at  their  original  resolution
(64~ms). The low-energy and  high-energy light-curves have been offset
by $-70$  and $-130$ counts/bin, relative to  their modelled location,
for clarity of presentation.}
\end{figure*}


\subsection{Burst profiles}
For bursts lasting  a few seconds or more it is  usually the case that
the light-curves contain several  peaks.  In a nanolensing model these
would be identified with caustic crossings, with each peak marking the
time  at which  the limb  of the  expanding source  first  touches the
caustic. Now  the relative timing offsets introduced  by parallax are,
of  course,  different  for  different  caustics, so  we  expect  that
parallax should alter  the time-interval between the peaks  of a given
burst, as  measured by physically separated  (but otherwise identical)
detectors.   The  effect is  evident  in  figure  9, which  shows  two
simulations  of an  expanding source,  as described  in  \S4.2, having
slightly different initial burst  locations. (In respect of the actual
calculations undertaken this is equivalent to displacing the observer,
and  thus simulates  the  effect  of parallax.)   One  of the  caustic
crossings  visible  in  figure 9  is  seen  at  almost the  same  time
($t\simeq0.013$) in  both light-curves, whereas  the other is  seen at
different  times   in  the  two  simulations.    The  expected  timing
difference is simply the quantity $\delta t_{12}$, estimated in \S5.1,
and  this  evaluates to  $|\delta  t_{12}|\sim1$~s, for  $z_s\simeq3$,
$b_{12}\sin\alpha\simeq3$~AU  and  $t\sim10$~s.   This value  is  much
larger  than the  limiting time-resolution  of the  Ulysses  and BATSE
instruments  (Hurley  et al  1992;  Fishman  et  al 1994)  and  should
therefore be  detectable if the  signal-to-noise ratio of the  data is
sufficiently high.   Such measurements offer  a very powerful  test of
the model we  are proposing, both because the  anticipated size of the
effect should render it measurable,  and because   {\it no other model
mimics this  behaviour.\/} In particular,
the test is  more powerful than one based  on apparent burst locations
(\S5.1),  because   there  is  no  requirement   for  absolute  timing
information;  only relative  timing is  important and  essentially all
spacecraft yield relative timing information with high accuracy.

To emphasise  the clarity of  this test let  us review what  should be
seen in the  absence of any parallax effects.  Two detectors observing
in  the the  same energy  band  should record  source flux  variations
which,  once the light-curves  have been  shifted so  as to  align the
times of burst onset, are in direct proportion to each other. Detector
efficiency   and  spacecraft   geometry  ordinarily   do   not  change
significantly  on  time-scales $\la100$~s,  so  a  single constant  of
proportionality  should hold  throughout each  burst. In  the  case of
spacecraft  in  low  Earth   orbit,  such  as  the  Compton  Gamma-Ray
Observatory, some fraction of the sky is occulted by the Earth, and it
is possible for ingress/egress to occur  during a burst. But this is a
rare phenomenon; moreover it is easily recognised, and therefore not a
concern in the present context. Particle backgrounds, and their trends
with time, inevitably differ between  any two detectors, but in normal
circumstances they also change  relatively little on the time-scale of
a  gamma-ray burst  and  a linear  trend  in the  background over  the
duration of the burst is  a suitable model for most purposes. Electron
precipitation events constitute a  variable particle background in low
Earth-orbit which in some respects can mimic a gamma-ray burst (Horack
et  al 1992);  however, these  events are  readily  distinguished from
gamma-ray  bursts   by  comparing   count  rates  amongst   the  BATSE
detectors.   Finally,  detector  dead-time   introduces  non-linearity
(different for different detectors) which can significantly affect the
recorded light-curve, causing the  count-rate to saturate if the burst
is bright. Saturation will not affect the peak location, because the
recorded count-rate is a monotonic function of the flux, but will
displace the burst centroid; however, saturation can be recognised
trivially, from the count rate, and affected bursts can be excluded
from consideration. In summary, none of the effects described above
should be confused with the parallax signal we have discussed.

In practice,  testing for parallax effects will {\it not\/} be done
with a pair of matched detectors, but with whatever instrumentation is
available in the IPN, so we will be comparing data from detectors with
different energy responses. This presents us with a potential problem
in that many of the ``pulses'' which are seen in GRB light-curves
exhibit spectral evolution (N96), and therefore have different shapes
in different energy bands. Consequently, with imperfect knowledge of
the (evolving) burst spectrum or the energy response of either of the
detectors, the pulse will appear to be slightly shifted in time, and
this effect can mimic a parallactic timing offset. Fortunately, not
all pulses are of this type (N96); a good fraction of pulses are
narrow, time-symmetric events, whose peak and centroid locations are
insensitive to photon energy (N96). This subset of pulses allows
parallax searches to be attempted even with detectors which are not
closely matched in their energy response. 

A  large   number  of  GRB   light-curves  are  freely   available  at
http://cossc.gsfc.nasa.gov/batse   (the  BATSE  archive)   with  64~ms
resolution in  each of  four energy bands.  For comparison  with these
data we  have searched the  literature for published  GRB light-curves
from  the Ulysses  spacecraft  (Hurley et  al  1992). This  comparison
offers  very large  baselines (up  to  6.3~AU), for  which the  timing
offsets  should  be   correspondingly  large.   Although  Ulysses  has
detected a  large number of  GRBs, including hundreds which  were also
detected by BATSE (Hurley et al  1999a,b), we were able to find only 3
published Ulysses  light-curves for GRBs in the  BATSE catalogue.  Two
of these  -- BATSE2151 (GRB930131, the ``Superbowl  Burst'', Hurley et
al 1994),  and BATSE143 (GRB910503;  Hurley 1992) -- have  very modest
projected  separations  between  the  spacecraft (1.4~AU  and  1.6~AU,
respectively). We  have therefore focussed our  attention on BATSE1663
(GRB920622),   for  which   the  Ulysses-BATSE   projected  separation
(i.e. transverse to  the line-of-sight to the burst)  was 3.8~AU. This
burst has been analysed in detail by Greiner et al (1995).

We digitised the Ulysses light-curve directly from figure 1 of Greiner
et al (1995) and this light-curve is reproduced in figure 10, together
with the  BATSE light-curve for the  same interval. The  BATSE data we
plot  here  are  the sum  of  the  two  lowest energy  channels  only,
corresponding to  the photon energy  range 25--100~keV; this  range is
similar to  the sensitive range of the  Ulysses detector (15--150~keV;
Hurley  et al  1992).  The BATSE  data  have had  a linear  background
subtracted from  them; the background  model was determined  by taking
the median values of  1~minute intervals starting 3~minutes before and
4~minutes after  burst onset.  (This  procedure could not be  used for
the Ulysses  data because  the published light-curve  does not  span a
large enough  time interval.)  The two light-curves  were then aligned
in  time  such  that  their  cross-correlation is  maximised,  and  an
estimate of the Ulysses detector background was made by maximising the
cross-correlation  as a  function of  assumed background,  at constant
temporal offset. (We note  that varying the assumed background yielded
insignificant  changes  in  the  temporal alignment.)   The  estimated
Ulysses background  was in  this way found  to be  107~counts/bin; the
peak value of the normalised cross-correlation between light-curves is
0.974.

Although  the  temporal  alignment  between  the  two  spacecraft  was
determined to a  precision of 10~ms, the accuracy  of the procedure is
poorer than this.  We  estimated the uncertainties associated with our
method  by cross-correlating  the Ulysses  data with  the  two highest
energy  BATSE  channels.  These  data  are for  the  same  burst,  but
represent photons  of energies higher  than those to which  Ulysses is
sensitive.  This  procedure yielded  an  alignment  which differed  by
70~ms,  and a  background rate  which differed  by  2~counts/bin, when
referred  to the  results  of the  cross-correlation  with the  lowest
energy  BATSE data. Our  method can  be reasonably  expected to  be in
error by less than the shifts exhibited here.

The flux scale in figure 10 is  chosen to be that of the Ulysses data,
and the BATSE count rates  have been rescaled accordingly (and we have
then added a constant background of 107~counts/bin to the scaled BATSE
data). This choice  is appropriate because the BATSE  data are of much
higher signal-to-noise  ratio than the  Ulysses data, and  the Ulysses
flux  scale is  thus germane  to the  statistical significance  of any
comparison between the two. We  have also rebinned the BATSE data from
the original value of 64~ms to the 250~ms of the Ulysses data, and for
clarity we have offset the BATSE data by $-100$~counts/bin.  It can be
seen from figure  10 that the BATSE and  Ulysses light-curves are very
similar, as  expected, although there some evident  differences in the
$t=16-19$~s region, and the Ulysses light-curve seems to manifest more
fluctuations  in the  region $t>20$~s  than would  be expected  on the
basis of  the BATSE  light-curve. Quantitatively we  can say  that the
BATSE  data form  an  acceptable model  of  the Ulysses  data, with  a
$\chi^2$ statistic of 76 for  the difference between the two, over the
first  19~seconds  of  the  burst, with  $(19/0.25)-3=73$  degrees  of
freedom.

However, the key question for us here is not the overall similarity of
the  two light-curves, but  whether or  not the  temporal substructure
occurs  at different times  at the  two spacecraft.   In figure  11 we
concentrate our  attention on the  central region of the  burst, where
the BATSE  light-curve manifests obvious  substructure.  In particular
the  regions $t=7.5-12$~s  and $t=14.5-19$~s  show a  number  of peaks
which  one can  use to  test the  model. We  can compute  the combined
$\chi^2$ of these  intervals: it is $\chi^2=48.5$, with  36 degrees of
freedom.  Clearly the BATSE-derived model  does not perform as well in
these  intervals  with temporal  substructure  as  it does  elsewhere;
however, the  model is still  acceptable, with this value  of $\chi^2$
being realised by chance in 8\% of trials. Moreover, a simple $\chi^2$
test does  not reveal whether  the model is performing  poorly because
the temporal substructure is shifted in time, or for some other reason
such as the different energy response of the Ulysses detector compared
to the combination of BATSE channels 1 and 2.

Examination of figure 11 suggests  that both of these effects may play
a role. In particular we note that the feature in the low-energy BATSE
light-curve just after $t=18$~s  is clearly chromatic, with a centroid
which occurs earlier  at higher energies, and we  must therefore allow
that in the Ulysses data  this feature may exhibit a profile different
from the BATSE model. By  contrast, at $t=15$~s, the BATSE model shows
a  well-defined  event,  for  which  we  can  tentatively  identify  a
counterpart  peak in  the  Ulysses light-curve,  but this  counterpart
occurs 0.4~s later.  Although there is some difference in the profiles
of this event between high- and low-energy BATSE light-curves, it does
not seem reasonable to  attribute the BATSE-Ulysses shift to chromatic
effects  because  the differences  are  modest  whereas  the shift  is
comparable to  the width of the  feature. Another case  where the data
suggest a  temporal offset occurs  around $t=11$~s.  Here we  find two
narrow, approximately time-symmetric peaks  in the BATSE model. In the
Ulysses data  there is  no question about  which peaks to  identify as
their counterparts -- the two unresolved peaks either side of $t=11$~s
-- but the second peak appears earlier in the Ulysses data than in the
BATSE  model.  It  does  not  seem reasonable  to  attribute  this  to
chromatic  effects because the  profiles of  the low-  and high-energy
BATSE data are quite similar. However, we still cannot be confident of
having found an example of the  effect we are looking for, because (i)
the  peak  is  so  narrow  that  it is  undersampled  in  the  Ulysses
light-curve, and (ii) the  modest signal-to-noise ratio of these peaks
in  the Ulysses  data mean  that  any comparison  cannot yield  highly
statistically significant results.

In  summary, we  have compared  published Ulysses  data  with archival
BATSE  data for  a single  burst having  a large  projected separation
between spacecraft.   This comparison yielded some  hints of parallax,
but the evidence is of  low statistical significance and therefore not
compelling.  By reprocessing  the raw  Ulysses  data --  which have  a
temporal resolution  eight times finer than  the published light-curve
-- it should be possible to  make a more meaningful comparison between
the two  datasets. In particular the higher  temporal resolution would
be  valuable  in studying  the  peaks  close  to $t=11$~s,  which  are
unresolved   in  the   published  Ulysses   light-curve.    A  further
improvement on  our analysis  would be to  derive a  model light-curve
using  the  point-by-point spectral  data  (``colours'') derived  from
BATSE,  together  with  the  spectral  response matrices  of  the  two
spacecraft.

\section{Discussion}
Although the  burst (BATSE1663)  studied in \S5.2  was chosen  for its
relatively large  projected separation  between the spacecraft,  it is
not extreme in this respect,  and there are $\sim80$ instances (Hurley
et  al 1999a,b)  of larger  projected separations  amongst  the bursts
detected by both BATSE and Ulysses. Indeed the principal criterion for
using the BATSE1663  data was that the data  are published. This burst
merited  detailed study,  and  hence publication  of the  light-curve,
because it happened to occur within the COMPTEL field-of-view (Greiner
et al 1995). Amongst the many other bursts in which one might consider
searching for  parallax, how is  one to choose? The  expected temporal
offset   is  large  (see   \S5.1)  if   (i)  the   projected  baseline
$b_{12}\sin\alpha$ between  spacecraft is  large, and (ii)  the source
redshift  is  low  (hence  $\varpi\simeq1$), and  (iii)  the  apparent
expansion  speed of  the source  ($\beta_{ap}$) is  low. Unfortunately
only the first  of these criteria can be  securely determined directly
from  the data  at hand.  However, if  all other  parameters  are held
fixed,  then we  expect  that  high expansion  speed  and high  source
redshift would both lead to  large numbers of caustic crossing events,
so it  makes sense to  select bursts which  exhibit a small  number of
peaks in  their light-curves. Ideally the overlap  amongst these peaks
should  be  small,  so  that  the  properties  of  each  peak  can  be
characterised more-or-less independently of the others. These criteria
also  help  to avoid  the  potential  problem  of peak  confusion:  if
parallax  displaces  a burst  peak  by  an  amount comparable  to  the
separation between peaks  in one of the light-curves,  then it becomes
difficult to  identify counterparts between the  light-curves from the
different  spacecraft.  Finally,  and  most obviously,  a  significant
detection  absolutely requires  high signal-to-noise  ratio,  so bright
bursts are strongly preferred.

If we could increase the baseline between the two detectors beyond the
few  AU  which  characterises  interplanetary  spacecraft,  our  model
predicts  that the amplitude  and location  of the  substructure peaks
would differ  more and  more as the  baseline increased, and  it would
eventually become difficult to  identify counterpart peaks between the
two  profiles.  Ultimately,  if the  separation were  to  reach values
$\ga60\sqrt{M_{-4}}$~AU,  the entire caustic  pattern would  differ as
seen from  the two locations,  and there would consequently  be little
resemblance between the recorded  burst profiles. Although there seems
to be no immediate prospect of  separating a pair of detectors by such
a large  distance, it is  nevertheless possible to reach  this regime,
and  thus to  observe completely  different profiles,  if  a gamma-ray
burst    is   ``macroscopically''    gravitationally    lensed.   More
specifically, if a gravitational  lens forms multiple images which are
split     by      an     angle     very      much     greater     than
$15\sqrt{M_{-4}}$~nano-arcseconds, then the caustic crossings seen  in
each  image will  bear little  resemblance to  one another.   This, of
course, would be true even for micro(arcsecond)lensing due to stars at
cosmological distances. Thus  in the case of lensing  by galaxies, and
aggregates  of  larger mass,  any  lensed  ``echo''  should exhibit  a
different  temporal profile to  that of  the counterpart  signal which
arrives first (cf. Williams  and Wijers 1997; Paczy\'nski 1986b; Blaes
and  Webster  1992). It  is  expected  that  gravitational lensing  by
galaxies should  yield echoes of roughly 0.1\%  of bursts (Paczy\'nski
1986b; Mao 1992), but this effect has never been clearly detected. The
null results to date (Nemiroff et  al 1994; Marani et al 1999) are not
highly  statistically  significant, nor  is  it  trivial to  recognise
echoes even  if they are simply  scaled copies of  the ``original'' --
see Wambsganss (1993), and Nowak  and Grossman (1994). This means that
the lack of  any identified echoes, while consistent  with the present
model, is not a decisive argument in its favour. On the other hand, if
examples  were  found with  complex  temporal substructure  faithfully
reproduced in an echo, then  it follows that the observed substructure
is not due to gravitational nanolensing.

It has  previously been noted that  if the Universe  is populated with
low mass lenses then interference phenomena may manifest themselves in
gamma-ray  burst data  (Gould  1992; Stanek,  Paczy\'nski and  Goodman
1993; see  also Deguchi  and Watson 1986).  The lens masses  for which
this  effect is  relevant are  usually thought  to be  many  orders of
magnitude smaller  than the planetary-mass bodies  considered here. In
the case  of lensing  by a  fold, where the  angular splitting  of the
images  vanishes  for  a  source  located precisely  on  the  caustic,
interference fringes  can be  realised for larger  masses than  in the
case  of isolated gravitational  lenses. However,  strong interference
fringes also require that the source  be at most comparable in size to
the Fresnel  scale -- $\sqrt{D_s\lambda}$ for  wavelength $\lambda$ --
otherwise  the fringe patterns  from different  regions of  the source
manifest  their maxima at  quite different  locations, and  the fringe
visibility (contrast) is much reduced. We have already estimated (\S5)
the  thickness  of  the  high  intensity  limb of  the  source  to  be
$\sim4\times10^{11}$~cm, so with $D_s\sim4$~Gpc we can expect interference
phenomena   to   be   important   only   for   $\lambda\ga10^{-5}$~cm,
i.e. longward of  the far ultraviolet region, for  the source model we
have employed.

It  is  worth  noting  that  the  presence of  a  caustic  network  on
nano-arcsecond scales has no immediate implications for the observable
properties of  the GRB  afterglow, because at  this late stage  in its
evolution  the observable  emission from  the blast-wave  extends over
angular  scales which  are  so  large that  the  net magnification  is
very close to the mean magnification. Microlensing  of the   afterglow
is, however, possible  (Loeb \& Perna 1998; Mao \& Loeb 2001), because
the  characteristic  angular  scale  of  the magnification  pattern is
much larger  in this case.

The  two  main discrepancies,  which  we  noted  in \S4,  between  the
modelled nanolensing variations and the observed temporal structure in
GRB light-curves  are the small  amplitude and long time-scale  of the
nanolensing fluctuations.  Given that the model parameters appropriate
to real sources  are typically not known with  any accuracy, one could
try to evade the second  of these discrepancies by considering sources
at  high redshift.  Specifically,  as the  source redshift  approaches
$z_s=10.6$, corresponding  to nanolensing optical  depth $\tau=1$, the
caustic  density  becomes  very  large  indeed,  and  the  variability
time-scale  correspondingly  decreases   (\S3.4).  Thus  the  observed
characteristic time-scale of order one second between ``pulses'' (N96)
matches the nanolensing variability  time-scale only for very low mass
lenses ($\sim10^{-8}\;{\rm  M_\odot}$) at  $z_s=1$, but for  sources at
$z_s\sim10$   the  implied   lens  mass   would  correspond   to  that
($\sim10^{-4}\;{\rm  M_\odot}$)  deduced  from  the  data  on  quasars
(Schneider 1993; Minty 2001).  This idea is, however, quickly disposed
of because as the angular scale of the caustic network shrinks so does
the width of the high  magnification regions close to the caustics. In
turn  this means  that a  source of  given size  will  exhibit smaller
variations as it crosses  the caustics, exacerbating the other notable
discrepancy  between  the  nanolensing  simulations and  the  observed
variability.  This aspect  of  gravitational lensing  at high  optical
depth,  where  the depth  of  modulation  vanishes for  non-point-like
sources  as  $\tau\rightarrow1$, is  well  known  (Deguchi and  Watson
1987).

We wish to draw attention to the fact that all of the results reported
in the present  paper are for shear-free environments,  and if a large
external beam shear is present  then the properties of any nanolensing
can  be quite  different.  High  shear is  expected  if, for  example,
nanolensing  occurs  close  to  a microlensing  caustic.  Calculations
appropriate to  this circumstance  will be reported  elsewhere (Walker
and Lewis 2003).

\section{Conclusions}
Motivated by an existing interpretation of quasar variability in terms
of  gravitational nanolensing,  we have  examined the  implications of
this model for  the observed properties of GRBs.  Using a self-similar
blast-wave model to represent the source (no intrinsic variability) we
find that the  light-curves of some of the  caustic crossings resemble
those of the  ``pulses'' commonly seen in GRBs,  and for high redshift
sources  the time-scale  of  the predicted  nanolensing variations  is
consistent  with  the  GRB  data.  However,  the  predicted  depth  of
nanolensing modulation is far too small to explain the deep variations
observed in GRBs,  and this problem is exacerbated if  the GRBs are at
high redshift. These results mean that the GRB data do not exclude the
nanolensing  interpretation  of  quasar variability;  conversely,  the
simplest (shear  free) nanolensing  model cannot explain  the observed
GRB variability. Despite this failure, there  are weak  indications in
the  published  InterPlanetary   Network  data  that  nanolensing  may
actually be responsible for some of the observed  variations  of GRBs:
the light-curves  for one GRB show  hints of parallax. This effect  is
uniquely associated with lens-induced variations, and thus motivates a
careful examination of existing IPN data.

\acknowledgements MAW  much appreciates  the hospitality of  the Raman
Research Institute, where this work was begun. We have benefitted greatly
from many helpful discussions with Dipankar Bhattacharya, Sunita Nair
and P.~Sreekumar, all of whom came within a hair's breadth of being
co-authors on the present paper. GFL would like to thank
Joachim Wambsganss for providing  his ray-tracing code and David Bowie
for his {\bf Low} album. We thank the referee, Robert Nemiroff, for
helpful comments.


\appendix

\section{Gamma-ray burst astrometry}
In this  Appendix we give details  of the astrometric  analysis of the
sample of bursts  referred to in \S5.1. The sample  was drawn from the
nine  bursts  observed  by   BATSE,  Ulysses  and  the  Mars  Observer
spacecraft (Laros et  al 1997), plus the 37  bursts observed by BATSE,
Ulysses  and Pioneer  Venus  Orbiter  (Laros et  al  1998). BATSE  was
capable of localising sources in its own right, by comparing the count
rates recorded  in different detectors,  each pointing in  a different
direction. This  was not, however,  a simple facility to  implement in
practice (Pendleton  et al 1999)  and the resulting  localisations are
only  accurate to  a  few degrees  ---  this point  has been  directly
verified by  comparing the  positions of solar  flares, as  located by
BATSE, with the position of  the Sun (Brock et al 1992).  Consequently
although  BATSE localisations  are  useful in  lifting the  degeneracy
between  the two intersection  points of  the IPN  loci, they  are not
accurate  enough  to reveal  systematic  errors  in the  triangulation
procedure  at the  arcminute level  we are  interested in.\footnote{At
this  point  we  should  draw  attention  to  the  anomalous  case  of
BATSE~2475, reported by Laros et al (1997), which apparently displayed
an  IPN  intersection  that  was significantly  discrepant  even  with
respect  to the  rather crude  BATSE localisation.   Having no  way of
understanding this result, Laros et al (1997) were obliged to conclude
that  Mars Observer had  not detected  the GRB  detected by  the other
spacecraft,  and had  in addition  responded to  another (unspecified)
event, which in  turn was not detected by the  other spacecraft in the
network, despite having the characteristics of a GRB. This coincidence
seems   rather   contrived.  An   alternative   explanation  for   the
timing/position  anomaly  of this  burst  can  be  found in  terms  of
parallax  (\S5  of  the   present  paper),  provided  that  BATSE~2475
exhibited  a  low  expansion  speed,  with  $\beta_{ap}\la40$.   (This
conclusion assumes a low  redshift burst, which nevertheless manifests
a  caustic crossing  event, and  makes use  of the  BATSE localisation
given  by Meegan et  al [1996].)  However, we  note that  the reported
discrepancy between BATSE and Mars Observer count rates for BATSE~2475
cannot  be so  readily explained.}  To  constrain any  such errors  we
therefore employed the localisations  which were obtained by the WATCH
experiment,  quite independently of  the triangulated  positions. This
experiment  employed  a   rotation  modulation  collimator  to  locate
sources, typically with  accuracies better than half a  degree (at the
$1\sigma$ level; Sazonov et al 1998).  The additional requirement that
each  burst have  been  observed by  WATCH  necessarily decreases  the
sample size and we are left with only nine bursts in our final sample;
these bursts and the  corresponding astrometric information are listed
in  table  A1,  while   the  localisations  themselves  are  presented
graphically in figure 12.

It is immediately apparent from  table A1 that the WATCH error circles
are too  large to determine  positional errors at  the level of  a few
arcminutes  (which  is  the  upper  end  of  the  range  expected  for
$\delta\alpha$ in the blast-wave model --- see \S5.1). However, as can
be seen from figure~12, there are a number of instances in this sample
where the  loci derived  from burst timing  intersect at a  very acute
angle,  and in  these  cases the  intersection  point of  the loci  is
displaced on the sky by an amount $\gg\delta\alpha$, so that errors of
the anticipated size might possibly be revealed by comparison with the
WATCH  localisations.  To  test this  possibility we  have  employed a
penalty,  $\Pi(\delta\alpha)$,  which is  a  function  of the  offset,
$\delta\alpha$, in the radius, $\alpha$, of one of the loci:
\begin{equation}
\Pi(\delta\alpha)=\left({{\delta\alpha}\over{\sigma_\alpha}}\right)^2
+ \left({{\xi}\over{\sigma_w}}\right)^2,
\end{equation}
where  $\xi=\xi(\delta\alpha)$ is the  angular separation  between the
intersection of the IPN loci and the WATCH location; $\sigma_w$ is the
estimated error  in the WATCH  location (one third of  the $3\,\sigma$
statistical error, plus the systematic  error, as quoted by Sazonov et
al  1998),   and  $\sigma_\alpha$  is  the  estimated   error  in  the
determination   of   $\alpha$    (one   third   of   the   $3\,\sigma$
statistical-plus-systematic  error   quoted  by  Laros   et  al  1997,
1998).  The  location  ($\delta\alpha_{min}$)  of the  minimum  value,
$\Pi_{min}$, of this  penalty function can, in some  cases, give us an
estimate of the value of $\delta\alpha$. One point to note is that the
WATCH error  distribution is ellipsoidal, but we  are approximating it
by a circular distribution.

Generally speaking, of  the three loci derived from  burst timing, one
(corresponding to the BATSE-Ulysses  baseline) has by far the smallest
errors and our  adopted procedure is therefore to  treat this locus as
an  absolute constraint  on the  burst location.  We then  discard the
annulus with the largest estimated  error, and use the remaining locus
to determine the  intersection point. The radius of  this second locus
is  then allowed to  vary by  an amount  $\delta\alpha$, over  a range
$\pm5\,\sigma_\alpha$, and for each value of $\delta\alpha$ we compute
the   separation,  $\xi$,  and   thus  the   value  of   the  penalty,
$\Pi(\delta\alpha)$. In general  the value of $\Pi_0\equiv\Pi(0)$ will
not   coincide    with   the   minimum   value    of   the   function,
$\Pi_{min}=\Pi(\delta\alpha_{min})$, and if $\Pi_0-\Pi_{min}\gg1$ then
it  is   clear  from   the  definition  of   $\Pi$  that   this  means
$\delta\alpha_{min}$   is  distinguishable   from  zero.    The  ratio
$\delta\alpha_{min}/\sigma_\alpha$ tells  us whether the  actual error
which we have detected is significantly larger than the expected error
in  $\alpha$. The results  of this  procedure are  given in  table A1,
where it can be seen that  there are four bursts (BATSE~451, 907, 1698
and   2387)  for   which   $\Pi_0$  is   significantly  greater   than
$\Pi_{min}$.  With the  possible exception  of BATSE~2387,  the errors
which we  measure in these cases (i.e.  $\delta\alpha_{min}$) are only
as   large  as   one  would   expect,  given   the   estimated  error,
$\sigma_\alpha$, in each case. Of  course it remains possible that the
systematic error we are looking for is in fact a major contribution to
the estimated value  of $\sigma_\alpha$, and we are  unable to exclude
this possibility with the information at hand.


\begin{table}
\scriptsize
\begin{center}

\caption{\centerline{Astrometry of gamma-ray burst sources}}
\begin{tabular}{crrrrrrrrrrr}
\tableline\tableline
\ \\
BATSE & R.A. & Dec. & $\alpha$\ & $3\,\sigma_\alpha$ & R.A. & Dec. &
$\sigma_w$\ & $\Pi_0$ & $\Pi_{min}$ & $\delta\alpha_{min}$ &
$|\delta\alpha_{min}|$ \\
Trigger \# & (deg) & (deg) & (deg) & (deg) & (deg) & (deg) & (deg) & & &
($\sigma_\alpha$)\ & (arcmin)\\
\ \\
\tableline
\ \\

451\tablenotemark{*} & 134.8255 & 18.4232 & 66.8162 & 0.0119 \\
    & 133.5243 & 18.7903 & 68.0884 & 0.0226 & 199.60 & --02.60 & 0.41 & 44
& 1.56 & 1.07 & 0.48\\

907\tablenotemark{*} & 342.4405 & --08.6911 & 44.1528 & 0.0641 \\
    & 342.9264 & --08.8461 & 44.5370 & 0.2785 & 297.37 & --04.71 & 0.56 &
944 & 3.03 & 1.09 & 6.0\\

1141 & 168.0347 & 06.4847 & 30.0358 & 0.0472 \\
     & 160.4516 & 09.2822 & 34.6366 & 0.2360 & 171.97 & --22.59 & 0.57 &
2.74 & 2.39 & --0.50 & 2.3\\

1473 & 155.1400 & 10.0896 & 51.2649 & 0.0096 \\
     & 157.1525 & 08.8837 & 51.0046 & 0.0100 & 132.25 & --36.39 & 0.22 &
0.09 & 0.08 & --0.07 & 0.01\\

1538 & 332.4865 & --09.1834 & 32.9477 & 0.0137\\
     & 338.9289 & --07.6437 & 33.7998 & 0.0189 & 323.07 & 22.53 & 0.29 &
1.01 & 1.00 & 0.10 & 0.04 \\

1698\tablenotemark{*} & 152.2635  & 05.1023   & 74.4660 & 0.0099\\
     & 161.6670  & --00.9283 & 63.3050 & 0.0164 & 221.43 & --30.75 & 0.26 &
7.14 & 4.19 & 0.32 & 0.11\\

1712 & 152.9064  & 04.6545 & 17.1004 & 0.2556 \\
     & 130.6922  & 19.5520 & 33.6130 & 0.3669 & 145.67 & --11.20 & 0.41 &
1.45 & 1.34 & 0.31 & 2.3\\

2387\tablenotemark{*} & 143.9743 & --11.1837 & 63.9320 & 0.0520 \\
     & 141.8929 & --23.8803 & 51.5670 & 0.6170 & 109.24 & --71.20 & 0.30 &
1195 & 5.75 & --2.24 & 28\\

2431 & 326.3289 & 11.8938 & 54.6810 & 0.0060 \\
     & 317.3416 & 23.2362 & 55.8180 & 0.0120 & 281.42 & --20.18 & 0.24 &
0.58 & 0.58 & 0.04 & 0.01\\

\ \\
\tableline
\end{tabular}
\tablenotetext{*}{The actual error in $\alpha$ can be gauged for this
burst, because
$\;\;\Pi_0-\Pi_{min}\gg1$.}
\tablecomments{Columns 2 \& 3 give the Right Ascension and Declination of the
centre of the IPN loci, with the corresponding dimensions of the annulus being
prescribed by the angular radius, $\alpha$ (column 4) and its associated error,
$\sigma_\alpha$ (column 5). Columns 6 \& 7 give the Right Ascension and
Declination
for the WATCH localisation, with the radius of the error circle, $\sigma_w$, in
column 8. All coordinates are J2000. Columns 9--12 give properties of the
penalty
function, $\Pi$ (equation A1), as deduced for each burst.}
\end{center}
\end{table}


\begin{figure*}
\plottwo{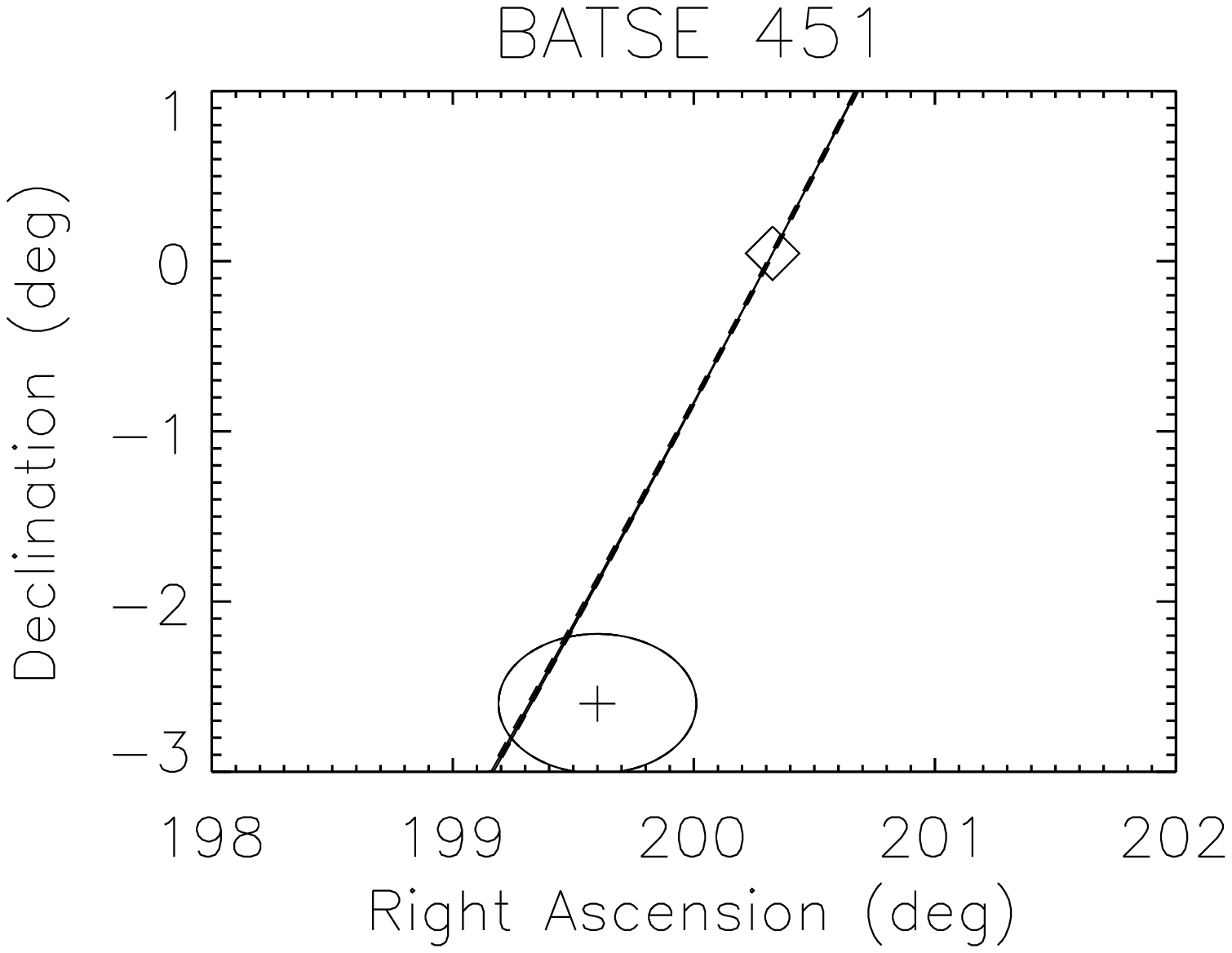}{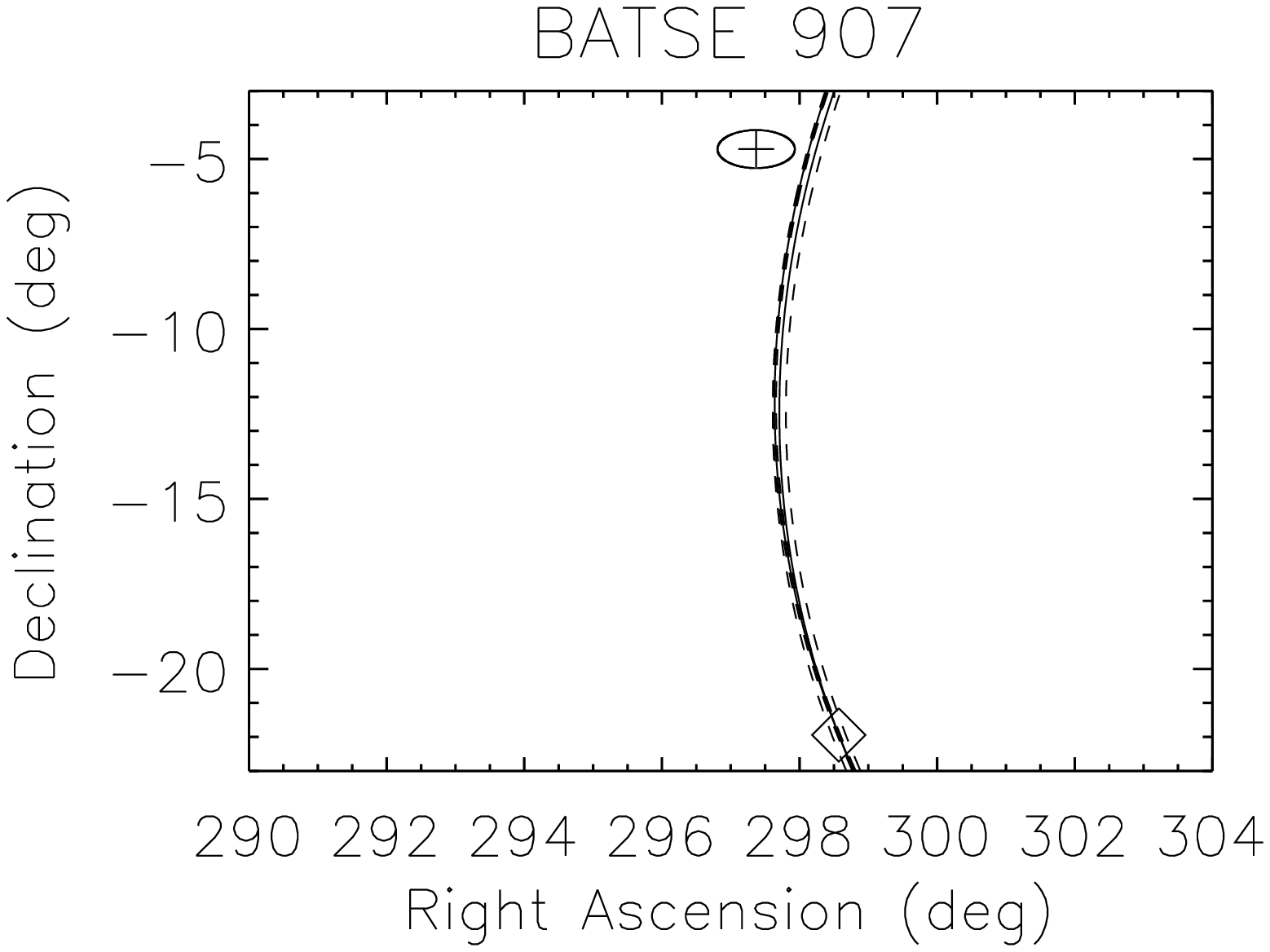}
\end{figure*}
\begin{figure*}
\plottwo{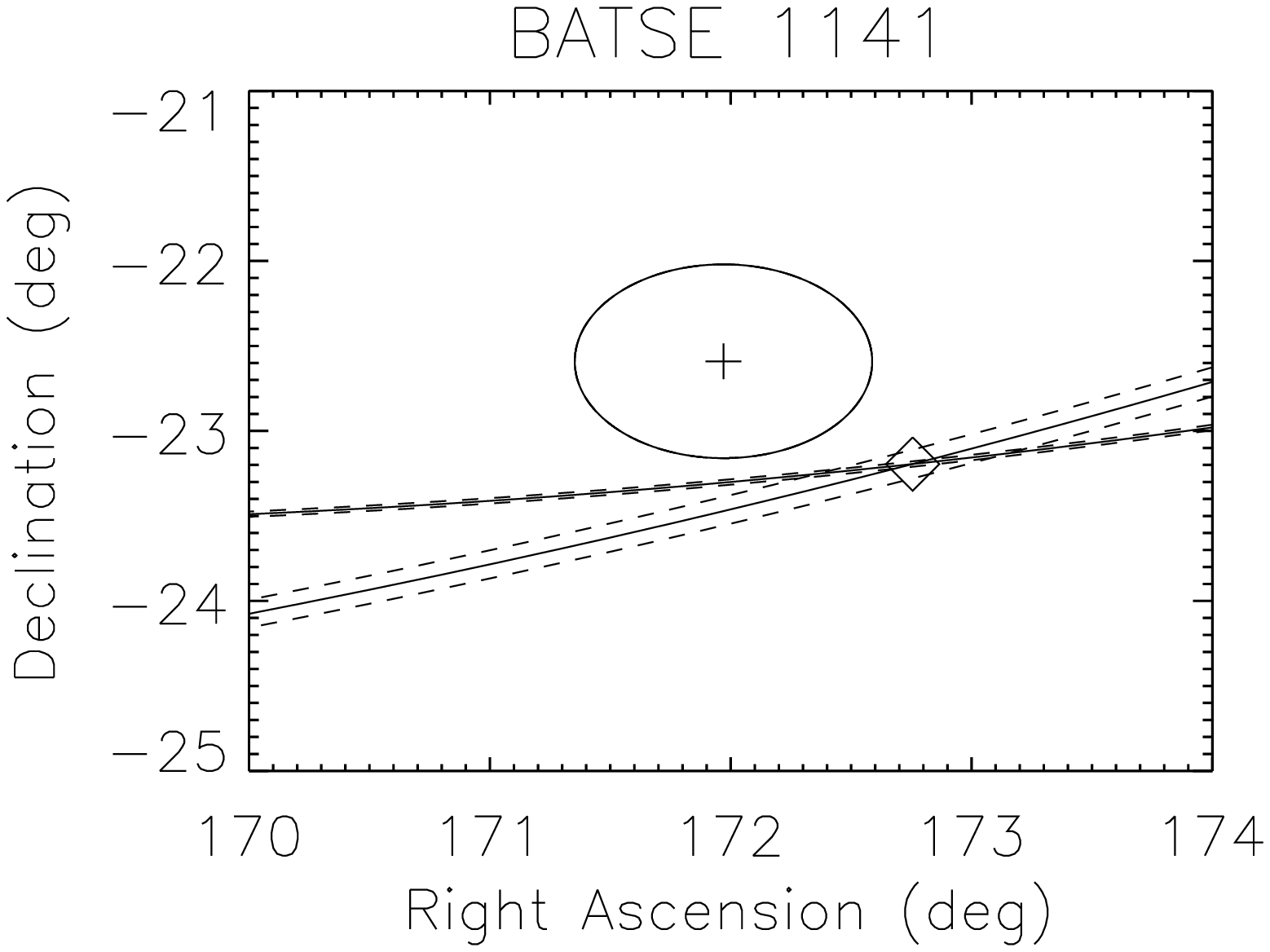}{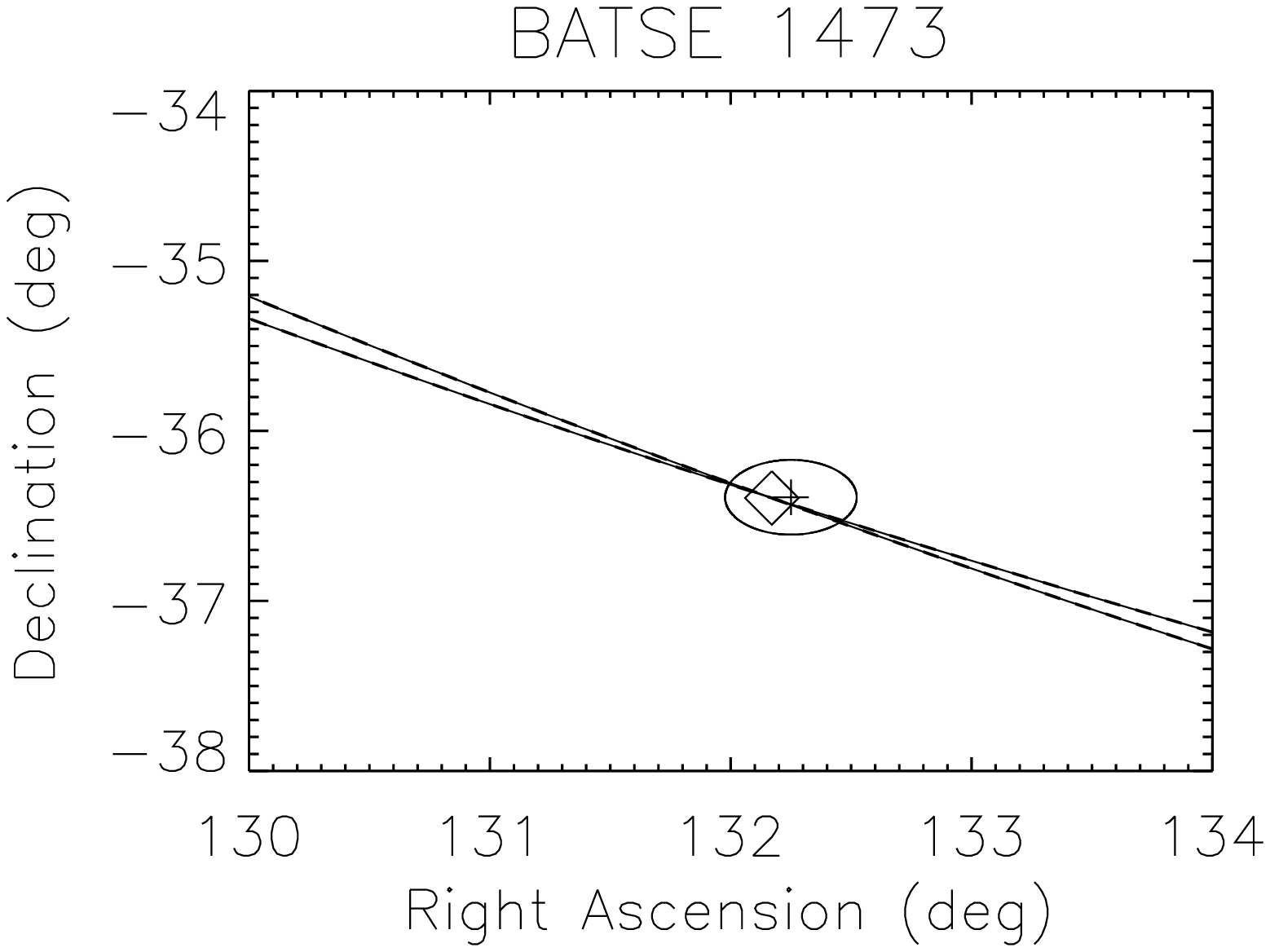}
\end{figure*}
\begin{figure*}
\plottwo{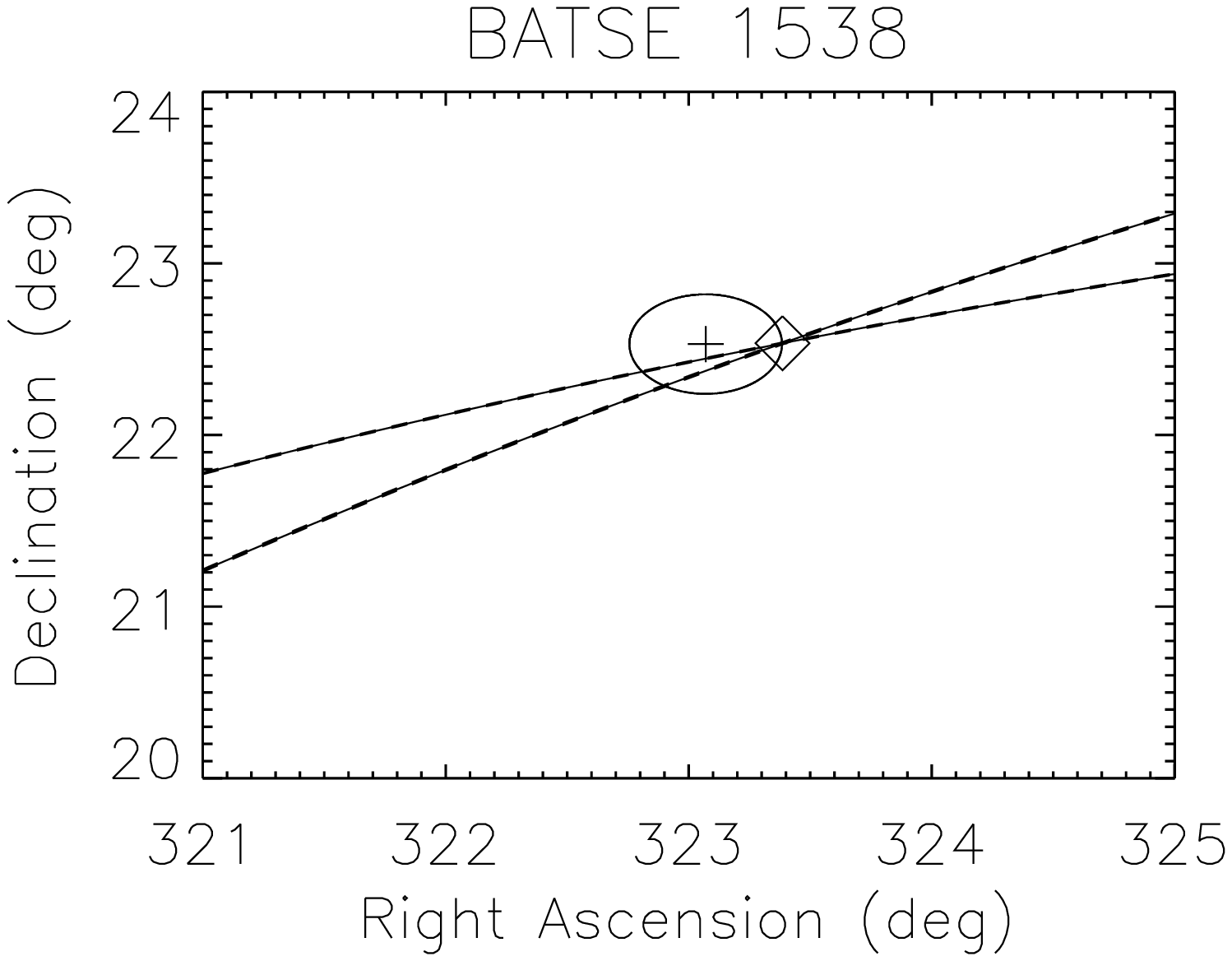}{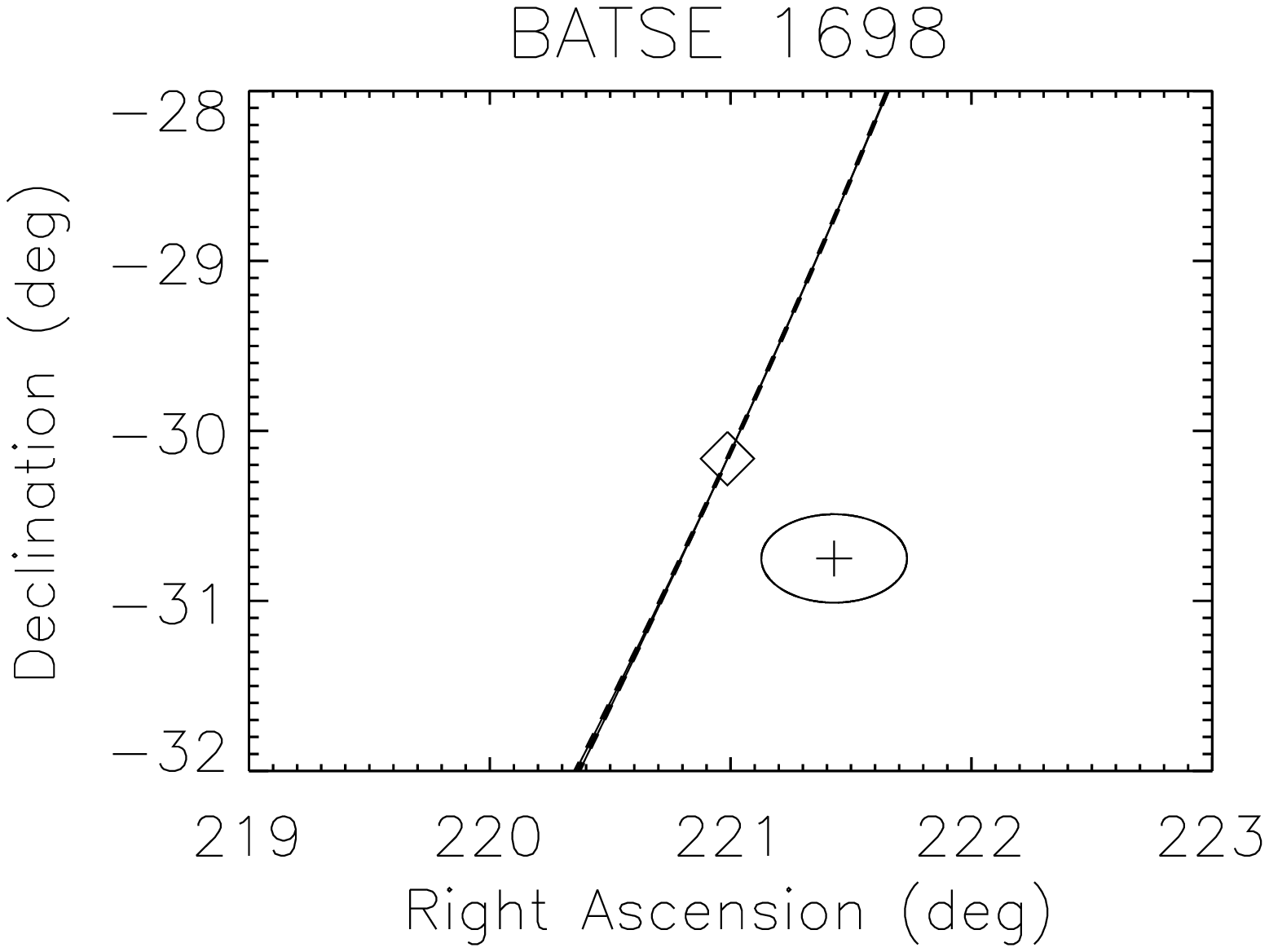}
\end{figure*}
\begin{figure*}
\plottwo{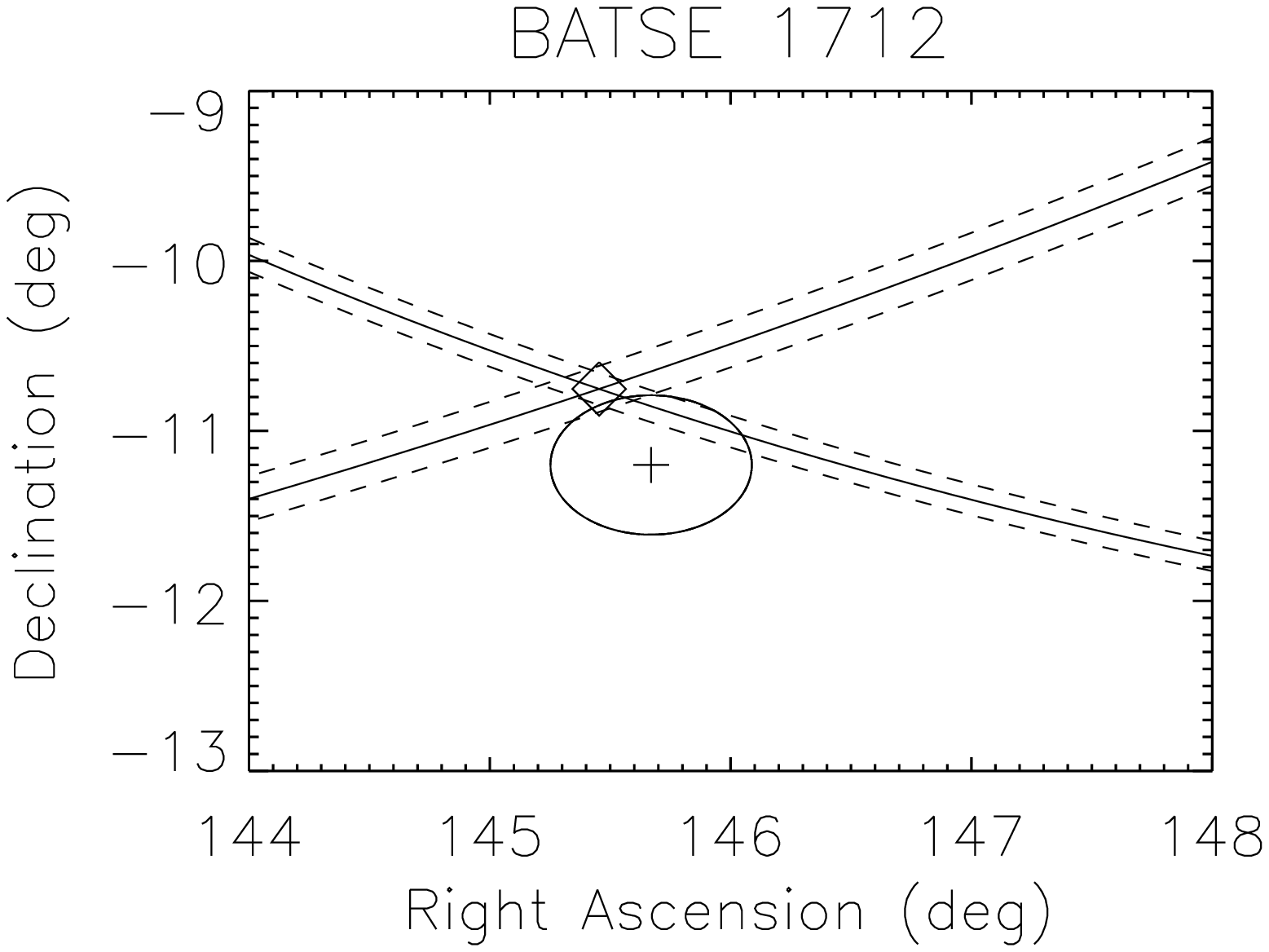}{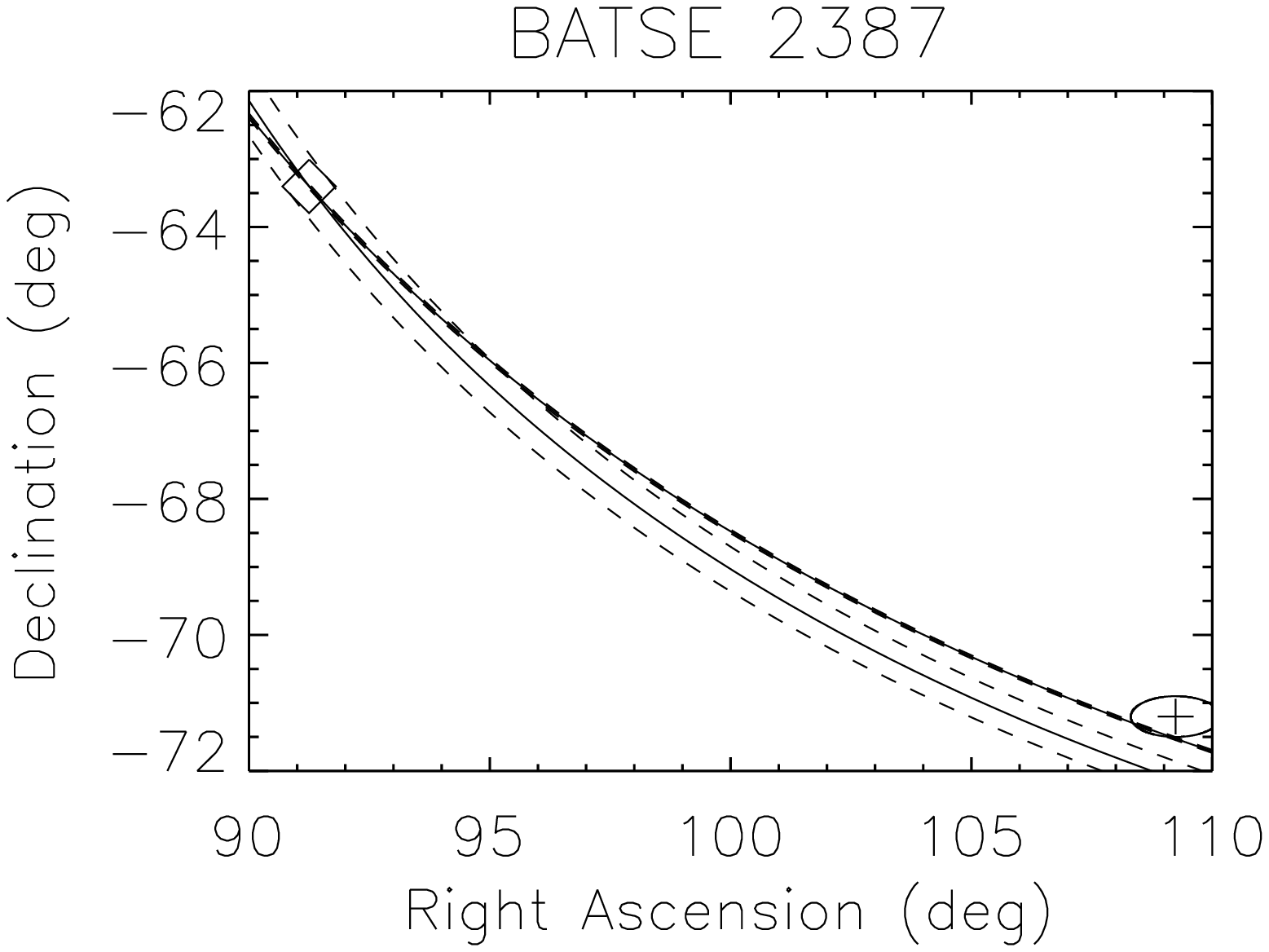}
\end{figure*}
\begin{figure*}
\plottwo{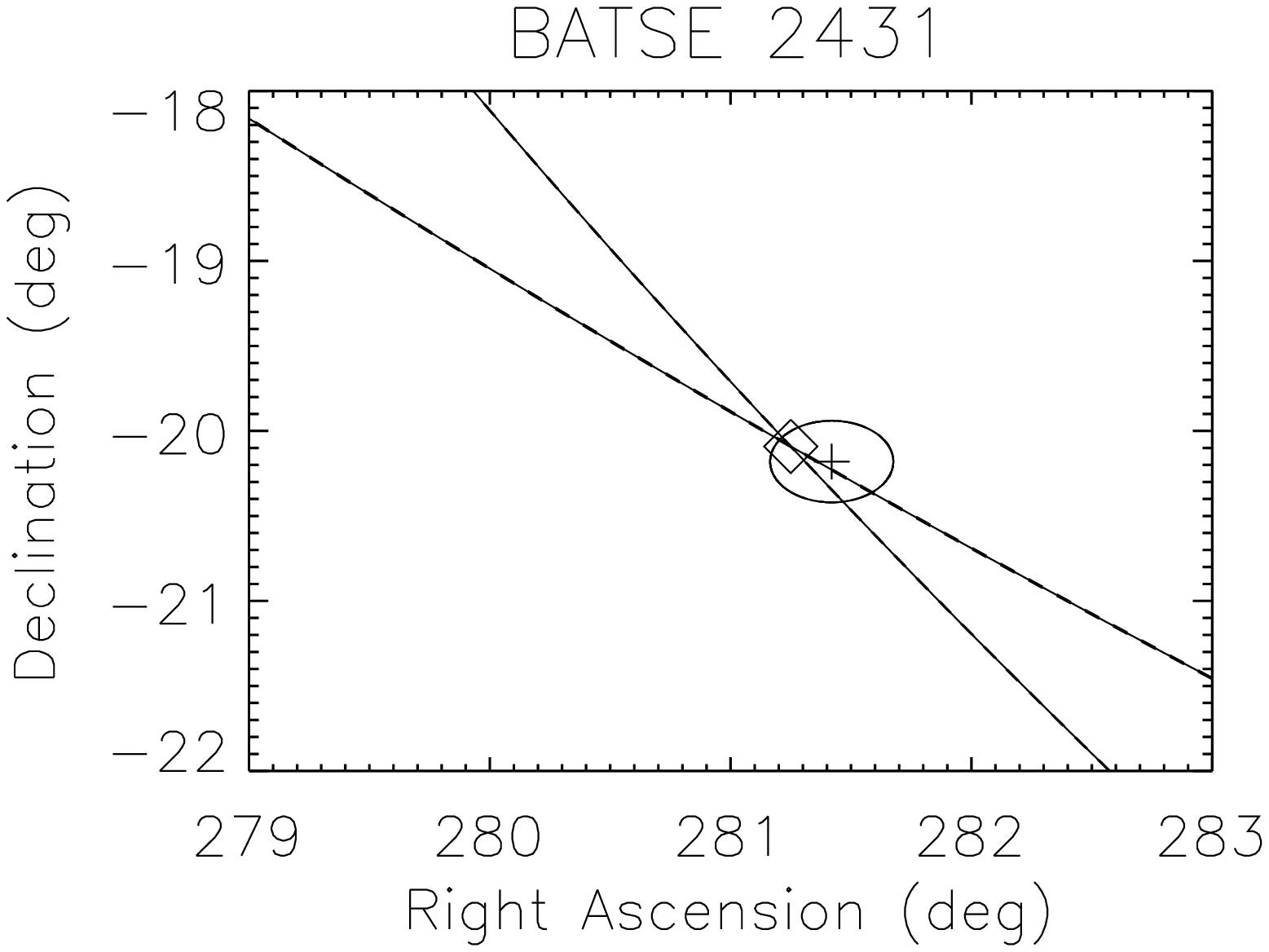}{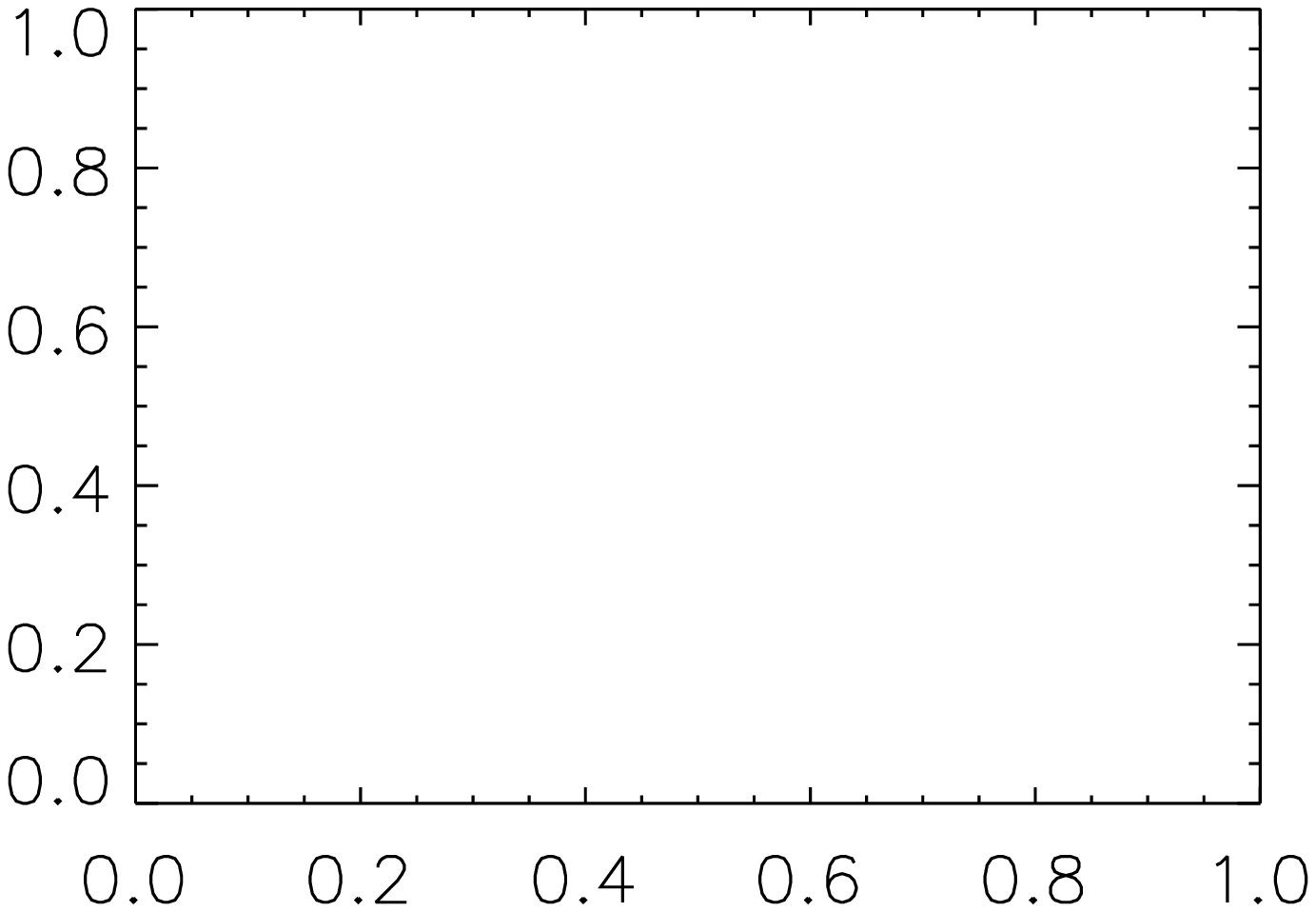}
\caption{Positions for a sample of 9 bursts which were located by both the
WATCH
instrument -- crosses, with $1\sigma$ (statistical plus systematic) error
circles
-- and the Inter-Planetary Network (Ulysses plus BATSE and either PVO or
MO). IPN loci
are shown (solid), with $\pm1\sigma$ errors (statistical plus systematic:
dashed). A
diamond marks the intersection of the IPN loci in each case.}
\end{figure*}

\end{document}